\documentclass[twocolumn]{aastex701}

\usepackage{amsmath}
\usepackage{soul}  

\begin{document}

\title{Modeling tails of escaping gas in exoplanet atmospheres with \texttt{Harmonica}}

\author[0000-0001-5097-9251]{Carlos Gascón}
\affiliation{Center for Astrophysics $\mid$ Harvard $\&$ Smithsonian, 60 Garden Street, Cambridge MA 02138, USA }
\affiliation{Institut d'Estudis Espacials de Catalunya (IEEC), 08860 Castelldefels, Barcelona, Spain}
\email{}

\author[0000-0003-3204-8183]{Mercedes López-Morales}
\affiliation{Space Telescope Science Institute, 3700 San Martin Drive, Baltimore MD 21218, USA }
\email{}

\author[0000-0003-2527-1475]{Shreyas Vissapragada}
\affiliation{Carnegie Science Observatories, 813 Santa Barbara Street, Pasadena CA 91101, USA}
\email{}

\author[0000-0002-1417-8024]{Morgan MacLeod}
\affiliation{Center for Astrophysics $\mid$ Harvard $\&$ Smithsonian, 60 Garden Street, Cambridge MA 02138, USA }
\email{}

\author[0000-0003-4328-3867]{Hannah R. Wakeford}
\affiliation{School of Physics, University of Bristol, H.H. Wills Physics Laboratory, Tyndall Avenue, Bristol BS8 1TL, UK}
\email{}

\author[0000-0000-0000-0000]{David Grant}
\affiliation{School of Physics, University of Bristol, H.H. Wills Physics Laboratory, Tyndall Avenue, Bristol BS8 1TL, UK}
\email{}

\author[0000-0002-6689-0312]{Ignasi Ribas}
\affiliation{Institut de Ci\`encies de l'Espai (ICE, CSIC), Campus UAB, c/ de Can Magrans s/n, 08193 Bellaterra, Barcelona, Spain}
\affiliation{Institut d'Estudis Espacials de Catalunya (IEEC), 08860 Castelldefels, Barcelona, Spain}
\email{}

\author[0000-0002-3645-5977]{Guillem Anglada-Escudé}
\affiliation{Institut de Ci\`encies de l'Espai (ICE, CSIC), Campus UAB, c/ de Can Magrans s/n, 08193 Bellaterra, Barcelona, Spain}
\affiliation{Institut d'Estudis Espacials de Catalunya (IEEC), 08860 Castelldefels, Barcelona, Spain}
\email{}

\begin{abstract}

Exoplanets that reside close to their host stars, and therefore receive substantial amounts of X-ray and ultraviolet radiation, are prone to suffer from strong atmospheric escape. This can lead to the creation of an envelope of escaping gas along the planet's orbital trajectory, often referred to as a tail. When transiting in front of their host star, these tails can not only produce larger depths in the transit light curves, but also introduce significant asymmetries between ingress and egress.  Using the publicly available software \texttt{Harmonica}, we present a method to model the light curves of transiting planets surrounded by extended envelopes of escaping gas, and subsequently infer the shape and size of the latter. We apply this method to the JWST NIRISS/SOSS observations of HAT-P-18b, which show pronounced helium tail features in its spectroscopic light curve of the metastable helium triplet at 10830 \r{A}. Our model reveals that, in order to fit the observed light curve of HAT-P-18b,  the planet must possess a trailing helium tail of $15.79^{+1.14}_{-1.05}$ planetary radii. We carry out injection-recovery tests to validate the effectiveness of the proposed methodology. We demonstrate that, with sufficient precision,  we would be able to fit a multi-layer envelope to the data, which would provide insight into the relative radial variations in the opacity profile.

\end{abstract}

\keywords{}

\section{Introduction} \label{sec:intro}

Atmospheric escape is now recognized as a key physical process driving the mass loss and long-term evolution of planetary atmospheres, playing a central role in shaping the demographics of short-period, highly irradiated exoplanets via photoevaporation and core-powered mass-loss \citep{owen2019, owen2024}.
The first observational probe of this phenomenon has traditionally been the Lyman-$\alpha$ line at 1216 \AA, which has provided evidence for hydrogen escape in multiple Hot Jupiter \citep[e.g.,][]{vidal2003, desetangs2014} and warm Neptune \citep[e.g.,][]{kulow2014, ehrenreich2014} exoplanets.  More recently, the metastable helium triplet line at 10830 \AA~has emerged as another probe of planetary outflows. While Hubble provided the first detection around the low-density Neptune-mass planet WASP-107b \citep{spake2018}, the high resolution achievable with ground-based spectrographs has greatly expanded the sample of planets with a detected  escaping helium envelope \citep[e.g.,][]{nortmann2018, Kirk2020, zhoujian2023}. 

Although its resolution is lower than that of ground-based observatories, the James Webb Space Telescope (JWST) is also beginning to contribute to helium escape studies thanks to its unprecedented spectrophotometric precision and its ability to capture long, uninterrupted light curves \citep{dossantos2023}.
One such example is the NIRISS/SOSS observation of the inflated Saturn-mass planet HAT-P-18b, which revealed not only a larger transit depth, but also a highly asymmetric light curve with a significant post-transit absorption spanning more than 2 hours \citep{Fu2022, fournier2024}.
As with HAT-P-18b, in multiple cases the helium excess absorption has been found to extend well beyond the planetary transit, suggesting the presence of extended leading and/or trailing helium tails spanning projected lengths of several planetary radii \citep{zhoujian2023, tyler2024, gully_santiago2024}. The number and variety of pre and/or post transit absorption in the light curves suggests a wide range of sizes and morphologies for the helium escaping tails, likely shaped by the star-planet environment and interactions.
Multiple theoretical and modeling efforts have been put into connecting our physical understanding with the excess absorption and asymmetry measured from transit spectroscopy \citep{bourrier2013, owen2014, dossantos2022, schreyer2024}, including 3D hydrodynamic simulations \citep{macleod2022, nail2025, macleod2025} and semi-analytical models \citep{ballabio2025}.

Here, we aim to address this problem from an empirical perspective. Without making any assumption on the physical processes taking place, what envelope properties (e.g., tail sizes, envelope shape and opacity...) can we constrain by fitting the transit light curve of a planet surrounded by an envelope of escaping gas? The shape and opacity profile of the outflow allow us to probe how the cumulative signal from the excess obscuration is spatially distributed. Accessing this multidimensional information opens the possibility of understanding both how material is lost from the planet itself (for example through concentrated tails, polar outflows, or isotropically) and how it might be shaped by its interaction with the stellar environment. Because there are likely multiple physical processes at play, a modeling effort and an extensive theoretical understanding is likely needed to fully interpret this multidimensional observational information we aim to extract. By making this sort of measurement we aim to add empirical evidence to those interpretations.

This letter is organized as follows: In Section \ref{sec:methods}, we define the formulation and methodology to parameterize and model the light curves of transiting exoplanets surrounded by  escaping gas envelopes. We perform injection-recovery tests in Section \ref{sec:injection} to test the performance and robustness of our method, and in Section \ref{sec:application} we apply the method to the recent JWST NIRISS/SOSS observations of HAT-P-18b. We present our conclusions and future work in Sections \ref{sec:conclusions} and \ref{sec:futurework}.

\begin{figure*}
    \centering
    \includegraphics[width=1.\linewidth]{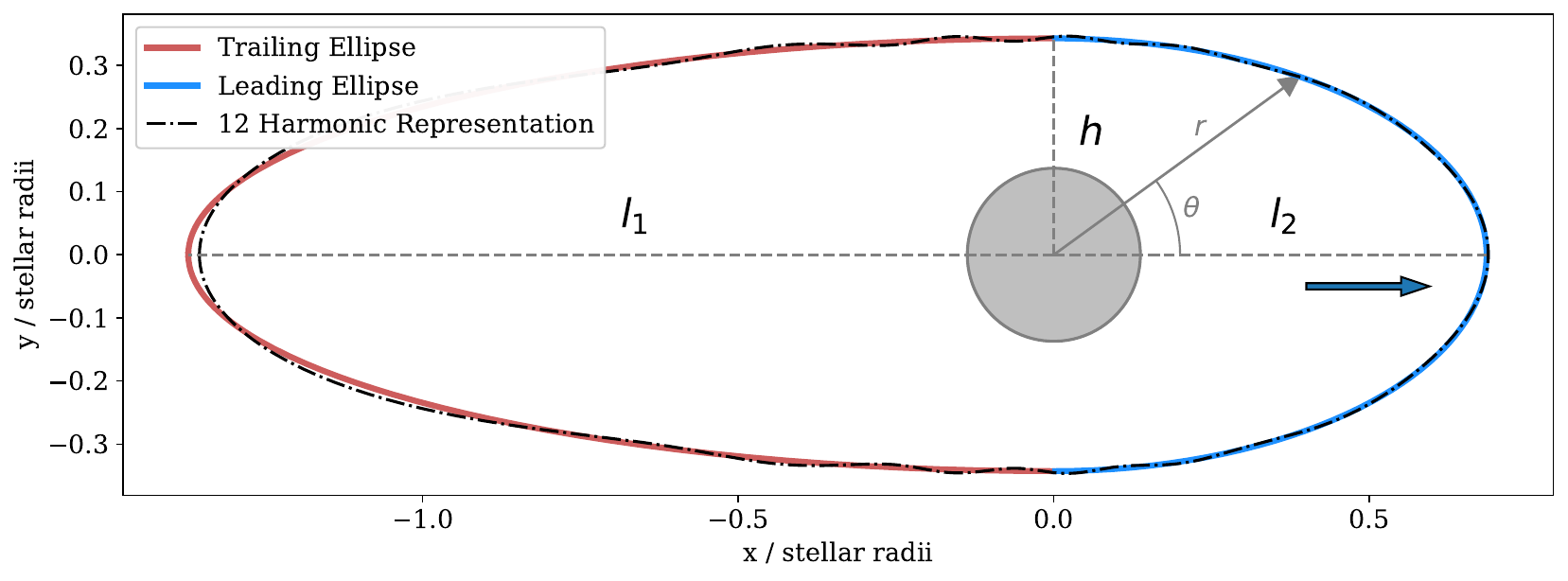}
    \caption{Schematic representation of the parametrization proposed in this study for an envelope of escaping gas around a planet (gray circle) moving in the direction of the blue arrow. The red semiellipse represents the trailing tail of length $l_1$, while the blue semiellipse represents the leading tail of length $l_2$. The two semiellipses share a common semi-major axis $h$, which describes the height of the envelope. The black dash-dotted line shows the 12 harmonic representation of the envelope.}
    \label{fig:diagram}
\end{figure*}

\section{Methodology} \label{sec:methods}

Considering the general case of a planet surrounded by an envelope of escaping gas both transiting its host star,  we first assume the planet's projection on the plane of the sky to be a perfectly opaque circle. The envelope of escaping gas, on the other hand, has an opacity $\alpha(r, \theta) \in [0, 1]$, which corresponds to the fraction of light blocked at any position ($r$, $\theta$) in the sky projected polar plane centered on the planet, where $r$ is the radial distance from center of the planet, and $\theta$ is the angle between the direction of $r$ and the direction of the leading tail as defined by the blue arrow in Figure \ref{fig:diagram}. The region for which  $\alpha(r, \theta) > 0$, which we assume to be continuous (i.e., with no gaps), defines the shape and size of the escaping gas envelope. We denote by $A_{\text{gas}}$ the total area covered by the gaseous envelope. The integral of the opacity over this area, which we refer to as effective area $A_{\text{eff}}$, can be written as
\begin{equation}
    A_{\text{eff}} = \int_{}^{A_{\text{gas}}}\alpha(r, \theta) r d\theta dr,
\end{equation}
This quantity has units of area, and gives us a sense of the amount of light being blocked by the gaseous envelope during transit. The shape, size and effective area of the envelope will imprint different features in the light curve when both the planet and the surrounding envelope transit in front of the star. Given a transit light curve, we aim to derive a simple parametrization that allows us to fit and evaluate some of these observables in a statistically robust manner. Below we present our formulation for the given problem.

\subsection{Modeling escaping tails}
We model the shape of the envelope of escaping gas around the planet as the combination of two semielipses with semimajor axes $l_1$ and $l_2$, and a shared semiminor axis $h$, as illustrated in Figure \ref{fig:diagram}. In polar coordinates, this shape can be written as
\begin{equation} \label{eq:ellipse}
    r_{\text{gas}}(\theta) = 
    \begin{cases}
        \frac{l_2h}{\sqrt{(l_2\sin{\theta})^2 + (h\cos{\theta})^2}}  & \text{if } -\pi/2 < \theta < \pi/2\\
        \frac{l_1h}{\sqrt{(l_1\sin{\theta})^2 + (h\cos{\theta})^2}} & \text{otherwise}
    \end{cases}  
\end{equation}

While this is a geometric simplification of what the real shape of the envelope might be, it provides sufficient degrees of freedom to consider different geometries: Assuming the planet moves in the direction indicated by the blue arrow in Figure \ref{fig:diagram}, $l_1$ and $l_2$ quantify the sizes of the trailing and leading tails, respectively, while $h$ quantifies the height of the envelope. In the simplest case, we assume the envelope to have a uniform opacity (i.e., $\alpha(r, \theta) = \alpha$), although we will also consider variable opacity later in this section. Denoting the transit of the planet and the gas envelope by $T_{\text{pl}}$ and $T_{\text{gas}}$, respectively, the combined light curve of a planet and its corresponding gas tail with uniform opacity $\alpha$ can then be expressed as 
\begin{equation} \label{eq:uniform_transit}
    T(t) = T_{\text{pl}} + \alpha(T_{\text{gas}} - T_{\text{pl}} ).
\end{equation}
In this case, the effective area corresponds to
\begin{equation}
    A_{\text{eff}} =  A_{\text{gas}}\alpha,
\end{equation}
where for our elliptical parametrization,  $A_{\text{gas}}$ would correspond to 
\begin{equation}
  A_{\text{gas}} = \frac{\pi l_1h + \pi l_2h}{2}  - \pi R_p^{2}
\end{equation}
Similarly, we can model an envelope with a non-uniform opacity by assuming that the latter is composed of $N$ layers, each described by a set of parameters ($l_{1, i}, l_{2, i}, h_{i}$), and with a different opacity $\alpha_i$, where $i \in {1,.., N}$. In this case, we assume the envelopes to be stratified into each other in a sorted manner, such that  $l_{1, i-1} < l_{1, i}$, $l_{2, i-1} < l_{2, i}$ and $h_{i-1} < h_{i}$, so the combined transit light curve can be written as
\begin{equation}\label{eq:multi_transit}
    T(t) = T_{\text{pl}} + \alpha_{1}(T_{\text{gas}, 1} - T_{\text{pl}}) + \sum_{i=2}^{N}\alpha_{i}(T_{\text{gas}, i} - T_{\text{gas}, i-1}),
\end{equation}
where $T_{\text{gas}, i}(l_{1, i}, l_{2, i}, h _{i})$ is the transit of the $i_\text{th}$ envelope. In this scenario, the effective area of the envelope can be written as
\begin{equation}
     A_{\text{eff}} = \alpha_1 A_{\text{gas}, 1} + \sum_{i=2}^{N}\alpha_{i}(A_{\text{gas}, i} - A_{\text{gas}, i-1}),
\end{equation}
where $A_{\text{gas}, i}$ is the projected area of the $i_\text{th}$ envelope.

\subsection{Calculating the transit light curve of the escaping tail with \texttt{Harmonica}} \label{sec:harmonica}

We use the publicly available software \texttt{Harmonica} \citep{grant2022transmission} to simultaneously model 
$T_{\text{pl}}$ and $T_{\text{gas}}$. \texttt{Harmonica} can compute the transit light curve of any shape that is expressed as the sum of harmonics, with the shape radius, $r$, at a given angle around the terminator $\theta$ given by
\begin{equation}
    r(\theta) = \sum_{n=0}^{N_c}a_n\cos{(n\theta)} + \sum_{n=1}^{N_c}b_n\sin{(n\theta)}, 
\end{equation}
where $a_n$ and $b_n$ are harmonic coefficients and $N_c$ is the number of harmonics (i.e., $2N_c + 1$ coefficients in total). As a result, we can take advantage of \texttt{Harmonica}'s capabilities to compute the transits as follows:
Given a set of envelope parameters $l_1$, $l_2$ and $h$, we compute the radial coordinates of the helium envelope via Equation \ref{eq:ellipse}, an subsequently perform a Fourier transform to find the corresponding harmonic coefficients $a_n(l_1, l_2, h)$ and $b_n(l_1, l_2, h)$ that better represent the helium envelope. For instance, the black dash-dotted line in Figure \ref{fig:diagram} shows the $N_c = 12$ harmonic representation of the helium envelope defined by the blue and red semiellipses. With the corresponding harmonic coefficients in place, we can then use \texttt{Harmonica} to compute the corresponding envelope transit light curve $T_{\text{gas}}$. Figure \ref{fig:examples} shows the transit light curve corresponding to different envelope morphologies. 

While \texttt{Harmonica} is built in a fast and computationally inexpensive manner, the increased number of harmonics needed to represent the slim elliptical shapes considered in this study can result in higher computing times. This can be particularly challenging in the multi-layer case presented in Section \ref{sec:multilayer}, for which a high number of iterations is needed. In that case, we speed up the fitting process by generating a grid of transits for a wide range of envelope parameters, and interpolating over the precomputed light curves during the MCMC run. 

\section{Injection-Recovery test} \label{sec:injection}

\begin{figure*}
    \centering
    \includegraphics[width=0.47\linewidth]{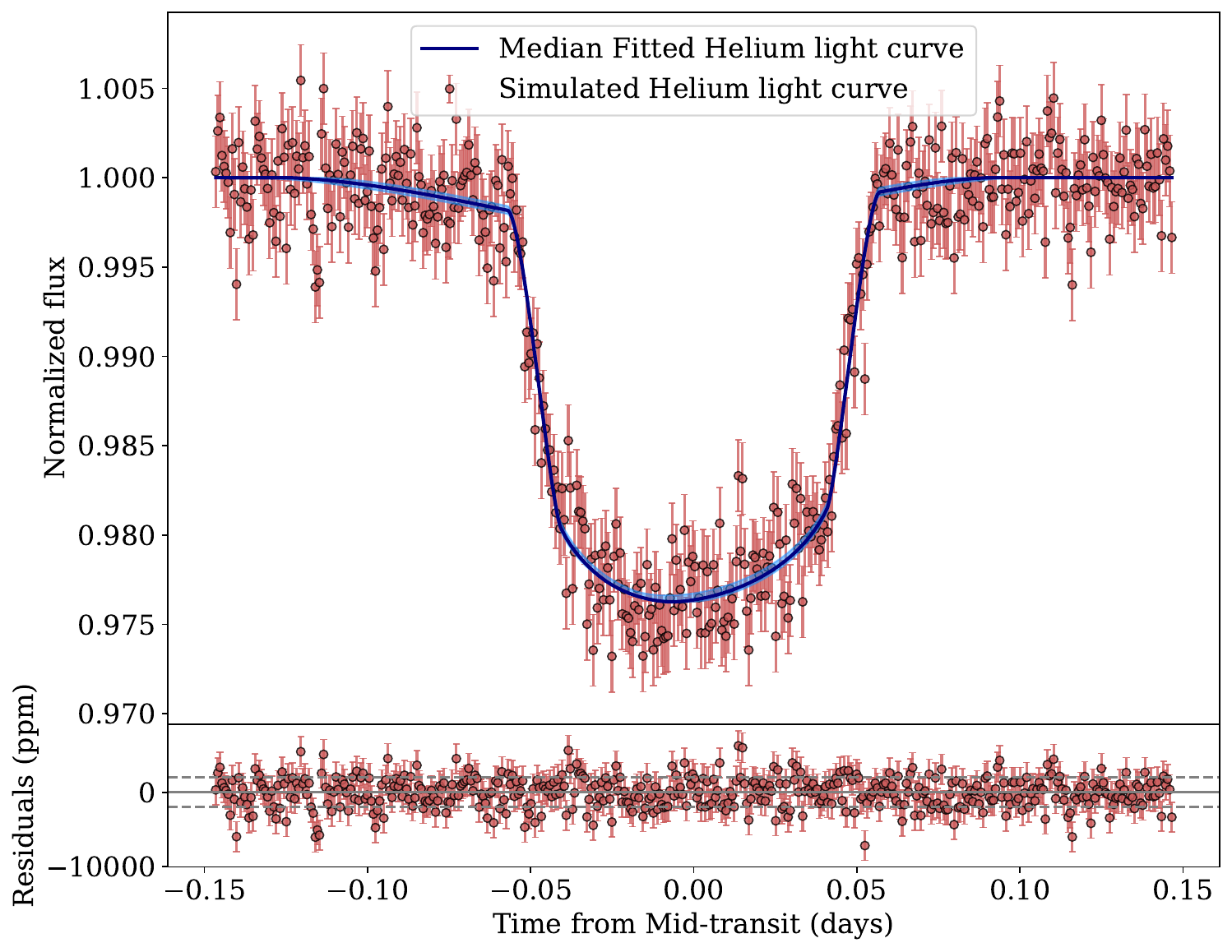} 
    \includegraphics[width=0.45\linewidth]{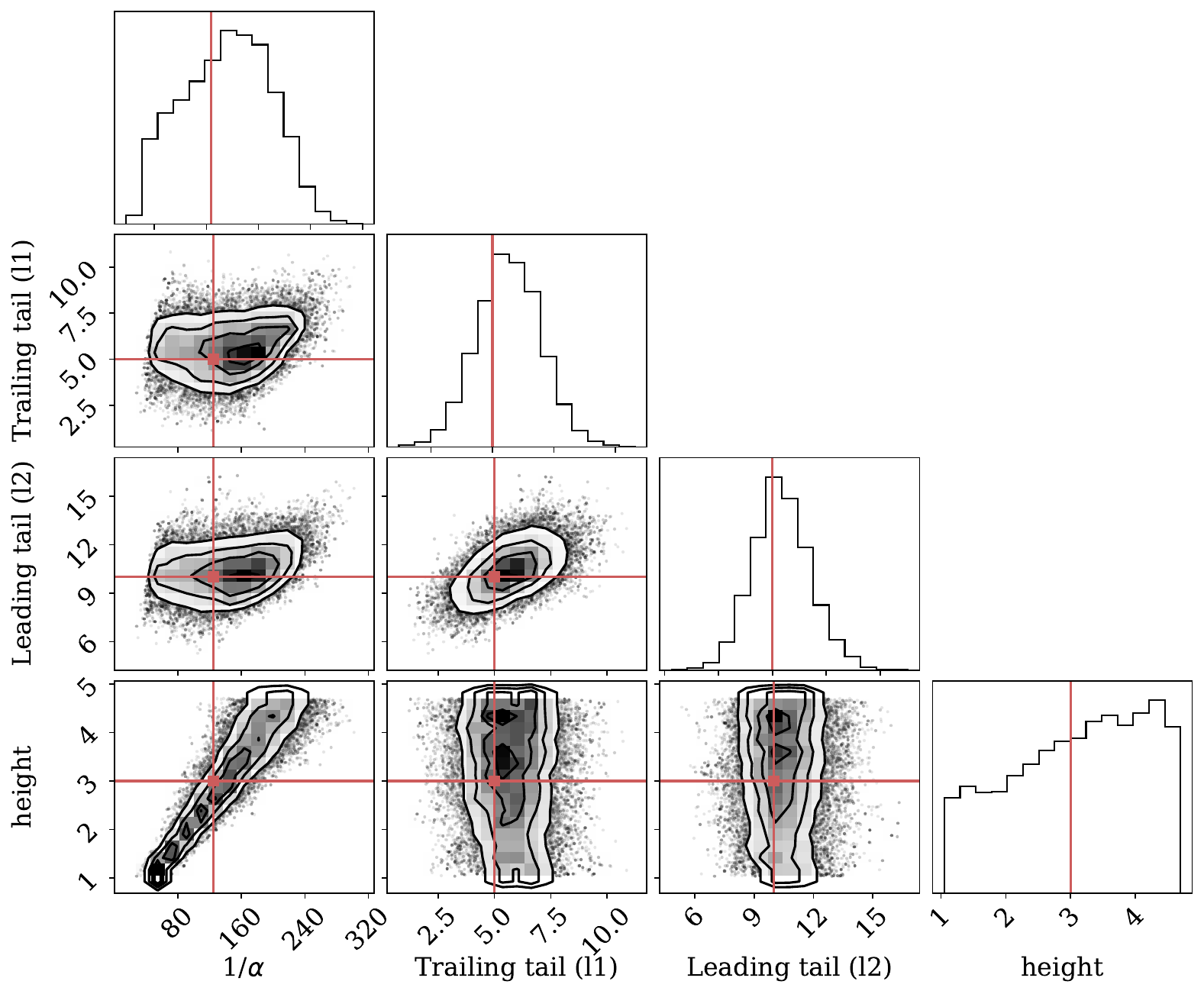}
    \includegraphics[width=1.\linewidth]{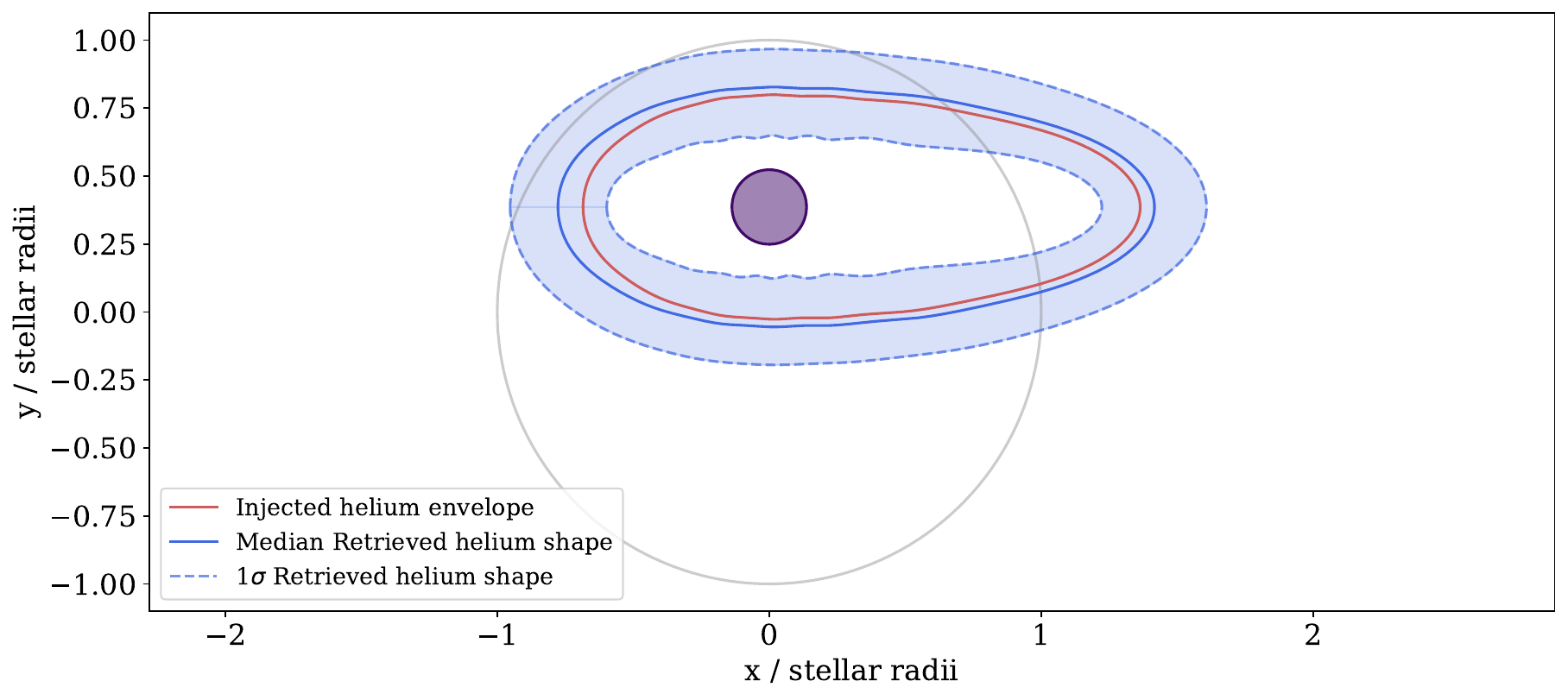}
    \caption{Main results from the injection-recovery test for the uniform envelope 2000 ppm case. The injected data is shown in red, and the recovered results are shown in blue. The top left panel shows the simulated light curve, together with the fitted model in blue. The top right panel shows the posteriors of the four fitted parameters, with the injected values shown in red. The bottom panel show the shape of the injected envelope in red, together with the median retrieved envelope in blue. The blue shaded region shows the $1\sigma$ uncertainty region, while the gray line represents the star and the purple circle illustrates the planet.}
    \label{fig:injection}
\end{figure*}

We assess the validity and robustness of the method described above by performing an injection-recovery analysis on synthetic data. For simplicity, we consider a system with the planetary and stellar parameters corresponding to the inflated Saturn-mass exoplanet HAT-P-18b, as it will be the target of the real data analysis presented in Section \ref{sec:application}. Furthermore, we adopt the duration and cadence of the NIRISS/SOSS observations of this same target, collected as part of the JWST ERO 2734 program, and adopt two light curve precision levels of 2000 ppm (similar to the real observations) and 200 ppm per data point to test the dependence of the results on the precision of the observations. All the system parameters referred to here are given in Table \ref{tab:sysparams}. The results of our tests are given in Tables \ref{tab:results} and \ref{tab:results_layers}. Below we describe the steps and results of the simulations for a uniform envelope and a multi-layer envelope.

\subsection{Uniform envelope simulation}

We first consider the simplest scenario of a planet surrounded by an envelope of escaping material with a uniform opacity $\alpha$. In particular, we simulate an envelope with a prominent leading tail $l_2 = 10R_p$, a smaller trailing tail $l_1 = 5R_p$, a height $h = 3R_p$ and an opacity $1/\alpha = 125$. Throughout the paper, we will assume that the envelope parameters are given in units of $R_p$. We also note that, in order to facilitate the fitting process, we will instead fit for and report the values of $1/\alpha$. To generate the synthetic observations, we first create a noiseless transit light curve for the planet and the helium envelope with \texttt{Harmonica}, as described in Section \ref{sec:methods}, and then combine both light curves together via Equation \ref{eq:uniform_transit}. We then add a Gaussian noise with a standard deviation corresponding to the two precision levels considered. For these tests, we do not include systematic noise, as we expect photon shot noise to generally be the dominant source of uncertainty, especially in JWST observations. The shape of the injected envelope, the corresponding synthetic transit light curve for the 2000 ppm per point case, and the input values for $l_1$, $l_2$, $h$, and $\alpha$ are shown in Figure \ref{fig:injection} in red. To fit the synthetic light curves, we fix all system parameters to the injected values, and fit only for the four parameters of the envelope ($l_1$, $l_2$, $h$, $\alpha$). We assume wide uniform priors for the four parameters, and only limit the height $h$ so that it is contained within the stellar radius. We perform the fits using the Markov Chain Monte Carlo (MCMC) sampler \texttt{emcee} \citep{emcee}, running 20 walkers, each with 400 steps as burn-in and 2000 steps for the production run. 

The main outputs of the MCMC run are presented in Figure \ref{fig:injection} in blue, where the posteriors, the retrieved helium envelope and the fit to the data are shown. The injected and recovered values for each parameter, together with the recovered values of the effective area $A_{\text{eff}}$ and the tail size ratio $l_1/l_2$,  are summarized in Table \ref{tab:results}. The results show that we are able to accurately recover the leading and trailing tail sizes, with precisions of $\sim1R_p$ in both cases.   We observe however a strong degeneracy between the envelope height, $h$, and the opacity, $\alpha$, evident by the correlation shown between the posterior distributions of the two parameters in the corner plot in Figure \ref{fig:injection}. 
This is because of the similar effect that both parameters imprint on the transit light curve, i.e. increasing (decreasing) the envelope height $h$ can be mimicked by decreasing (increasing) the opacity $\alpha$. We explicitly demonstrate this in Figure \ref{fig:degeneracy}, in Appendix \ref{sec:app_ex}. While we still get an estimate on the opacity which is in good agreement with the injected value (Table \ref{tab:results}), the degeneracy significantly enlarges the errorbars and therefore precludes us from accurately constraining the opacity and the height (and hence the aspect ratio) of the envelope simultaneously. Despite this degeneracy, we are however able to constrain the effective area $A_{\text{eff}} = 0.520^{+0.041}_{-0.039} R_{p}^2$ , which is in good agreement with the injected value of 0.541 $R_{p}^2$. The effective area essentially combines both parameters into one single metric and still provides meaningful physical insight. For instance, if we are able to better constrain the opacity from theoretical models or simulations, we can in turn infer the height $h$ of the outflow, and vice versa. We repeat the same procedure for the case in which the light curve precision is 200 ppm, with the fitted values shown in Table \ref{tab:results}. In this case, we are able to recover the leading and trailing sizes with precisions of $\sim 0.1 R_p$. While the degeneracy between $h$ and $\alpha$ still persists, evident by the large errorbars obtained for both parameters, the effective area is again much better constrained, with relative precisions similar to those achieved for the tail sizes.  

\begin{deluxetable*}{cc|c|ccc}
\tablecaption{Retrieved envelope parameters for the injection-recovery test and for the NIRIS/SOSS observations of HAT-P-18b.}
\label{tab:results}
\tablehead{\colhead{Variable} & \colhead{Units} & \colhead{HAT-P-18b} & \colhead{Injected} & \colhead{Recovered} & \colhead{Recovered} \\
\colhead{} & \colhead{} & \colhead{} & \colhead{} & \colhead{(2000 ppm)} & \colhead{(200 ppm)}}
\startdata
$l_1 $ & $R_p$ & $15.79^{+1.14}_{-1.05}$ & 5 & $5.67^{+1.28}_{-1.29}$ & $5.03^{+0.12}_{-0.11}$ \\
$l_2$  & $R_p$ & $5.11^{+0.99}_{-0.89}$ & 10 & $10.38^{+1.37}_{-1.24}$  & $9.99^{+0.14}_{-0.14}$ \\
$h$  & $R_p$ & $3.60^{+0.78}_{-1.05}$ &  3 &  $3.21^{+1.02}_{-1.33}$& $2.77^{+0.86}_{-0.69}$ \\
$1/\alpha$  & - & $206^{+40}_{-58}$ & 125 &  $145^{+49}_{-59}$ &  $116^{+35}_{-30}$ \\
$l_1/l_2$  & - & $3.08^{+0.60}_{-0.45} $ & 0.5 & $0.54^{+0.10}_{-0.10}$ & $0.50^{+0.01}_{-0.01}$ \\
$A_{\text{eff}}$  & $R_p^2$ & $0.562^{+0.036}_{-0.035}$ &  $0.541$ & $0.520^{+0.041}_{-0.039}$ & $0.536^{+0.014}_{-0.009}$ \\
\enddata
\end{deluxetable*}

\subsection{Multi-layer envelope simulation}\label{sec:multilayer}

\begin{figure*}
    \centering
    \includegraphics[width=1\linewidth]{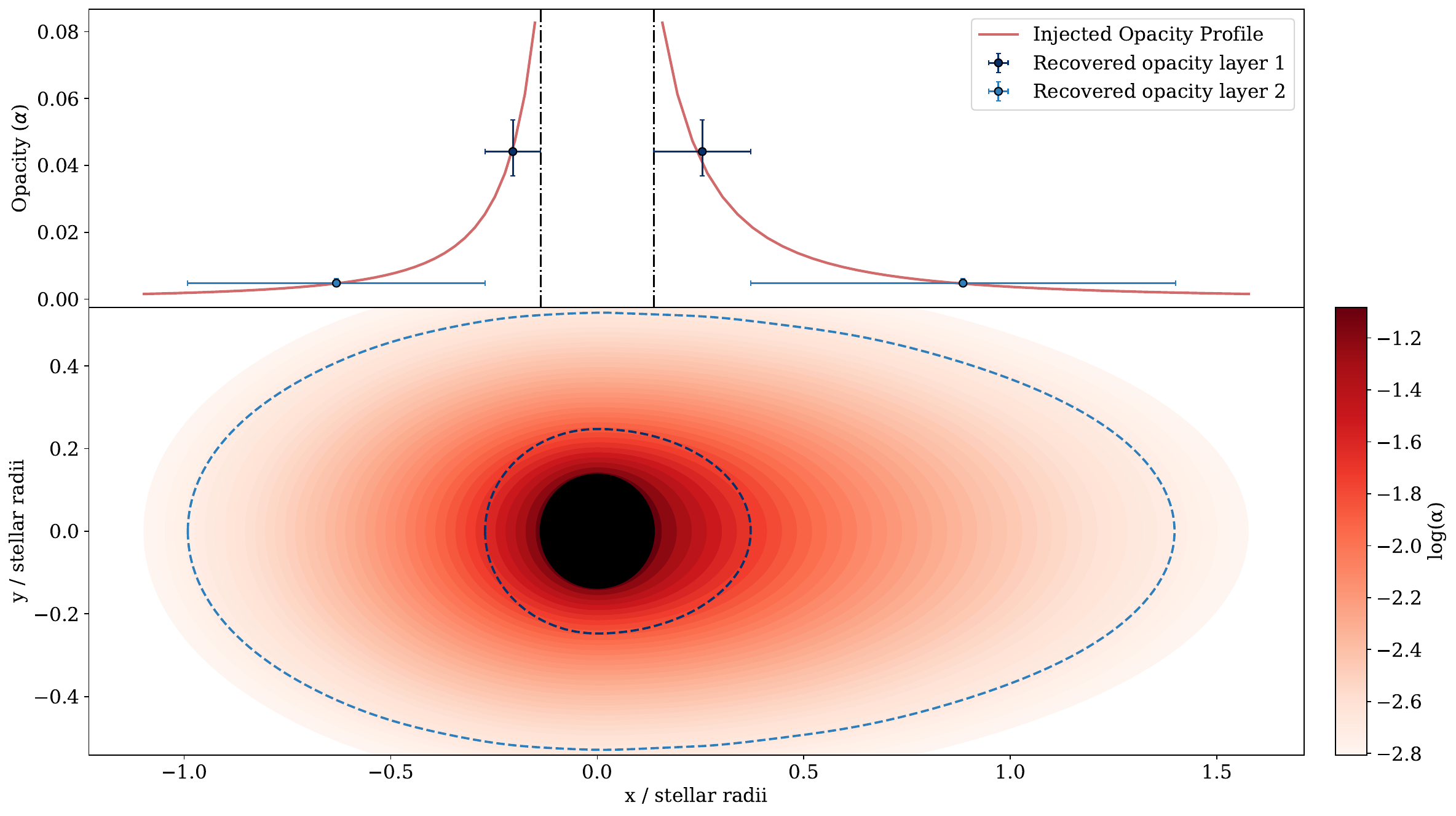}
    \caption{Results from the injection-recovery test for the helium envelope with a variable opacity. The top panel shows the inputted absorbance profiles (red curves), together with the empirically-retrieved absorbance profiles (red points with error bars) for the 2 layer envelope 200 ppm case. The bottom panel shows the injected variable envelope in different shadings of red, with the gray dashed lines indicating the position and shape of the fitted layers.}
    \label{fig:variable_envelope}
\end{figure*}

We also consider a more realistic scenario in which the planet is surrounded by an envelope of escaping material with variable opacity $\alpha(r, \theta)$. As described in Section \ref{sec:methods}, we approximate a variable envelope with a sum over $N$ nested shells (or layers) of decreasing opacity
(i.e., if $r_2 > r_1$ $\Rightarrow$ $\alpha(r_2, \theta) < \alpha(r_1, \theta)$ ).  
Each shell $i$ is defined by the envelope parameters ($l_{1, i}$, $l_{2, i}$, $h_{i}$), such that
\begin{eqnarray}
    l_{1, i} = 1 + f_{i} \nonumber\\
    l_{2, i} = 1 + c_{1}f_{i}\\
    h_{i} = 1 + c_{2}f_{i}\nonumber\, 
\end{eqnarray}
where $f_i$ corresponds to the scale parameter and everything is expressed in units of $R_p$. Here, $c_1$ and $c_2$ are constants defining the aspect ratio and tail asymmetry of the shells. For this example, we assume $c_1 = 3/2$ and $c_2 = 1/2$, and use a total of $N=40$ shells to simulate the envelope. The opacity of a given shell $i$ is given by an inverse decay profile of the form $\alpha_i = \frac{a}{(f_i + 1)^n}$. For this particular case we adopt $a = 0.1$ and $n = 2$, corresponding to a constant velocity gas expansion. The bottom panel in Figure \ref{fig:variable_envelope} illustrates the injected variable envelope, with the opacity profile represented by the red solid lines in the upper panel. The simulated transit light curve for the 200 ppm case, computed via Equation \ref{eq:multi_transit}, is shown in the top panel of Figure \ref{fig:multi_layer_results}. 

We fit the simulated light curves and recover the injected profile with \texttt{Harmonica} by fitting $M$ layers, each with a shape defined by the three variables $l_{1, j}$, $l_{2, j}$, $h_{j}$, and with an opacity $\alpha_j$, where $j \in {1, ..., M}$. We assume no prior information, but instead allow the four parameters describing  each layer to vary freely with uniform priors, constrained only by the relation of inclusion within each other. As a result, we fit a total of $4M$ parameters. For the 2000 ppm case we try to fit only for one and two layers, while for the 200 ppm case we run $M \in \{1, 2, 3, 4\}$. The results of these fits are summarized in Table \ref{tab:results_layers}, where for each case we report the derived values of A$_{\text{eff}}$ and the Akaike Information Criterion (AIC), which we use to evaluate the most suitable number of layers for each of the precisions considered.
For the 2000 ppm case, the best fit based on the AIC is provided by the one layer envelope (i.e., uniform envelope). We observe how, even if the injected envelope has a variable opacity, we are still able to get a good estimate of the effective area (i.e., the amount of light that is being blocked by the escaping gas) from a simple uniform (single-layer) envelope. The two layer envelope, on the other hand,  has a significantly lower AIC, meaning that one single uniform layer is enough to describe the helium envelope given the precision of the observations. 

When we increase the precision to 200 ppm, however, the single envelope is not the best fit to the data. In this case, the two-layer envelope provides the lowest AIC (Table \ref{tab:results_layers}), with an effective area in better agreement with the injected value. 
The blue points in the top panel of Figure \ref{fig:variable_envelope} show the recovered layers for the $M=2$ case, with the vertical errorbars showcasing the uncertainty in $\alpha$, and the horizontal errorbars symbolizing the radial extent of the layer. Figure \ref{fig:multi_layer_results} in Appendix \ref{sec:app_inj} shows additional information corresponding to the $M=2$ case, including the posterior distribution of the fitted parameters. Although some degeneracy still persists, we observe that the opacity of both layers, especially the inner layer, is relatively well constrained and in good agreement with the injected profile (Figure \ref{fig:variable_envelope}). This suggests that, with sufficient precision, the multi-layer analysis is able to provide insights into the relative radial variations in the opacity profile. As we increase the number of layers to $M=3$ and $M=4$, the effective area still agrees with the injected value, but the AIC decreases as a result of the increased number of variables introduced in the fit.

\section{Application to HAT-P-18\lowercase{b's} Helium Escape} \label{sec:application}

\begin{figure*}
    \centering
    \includegraphics[width=0.495\linewidth]{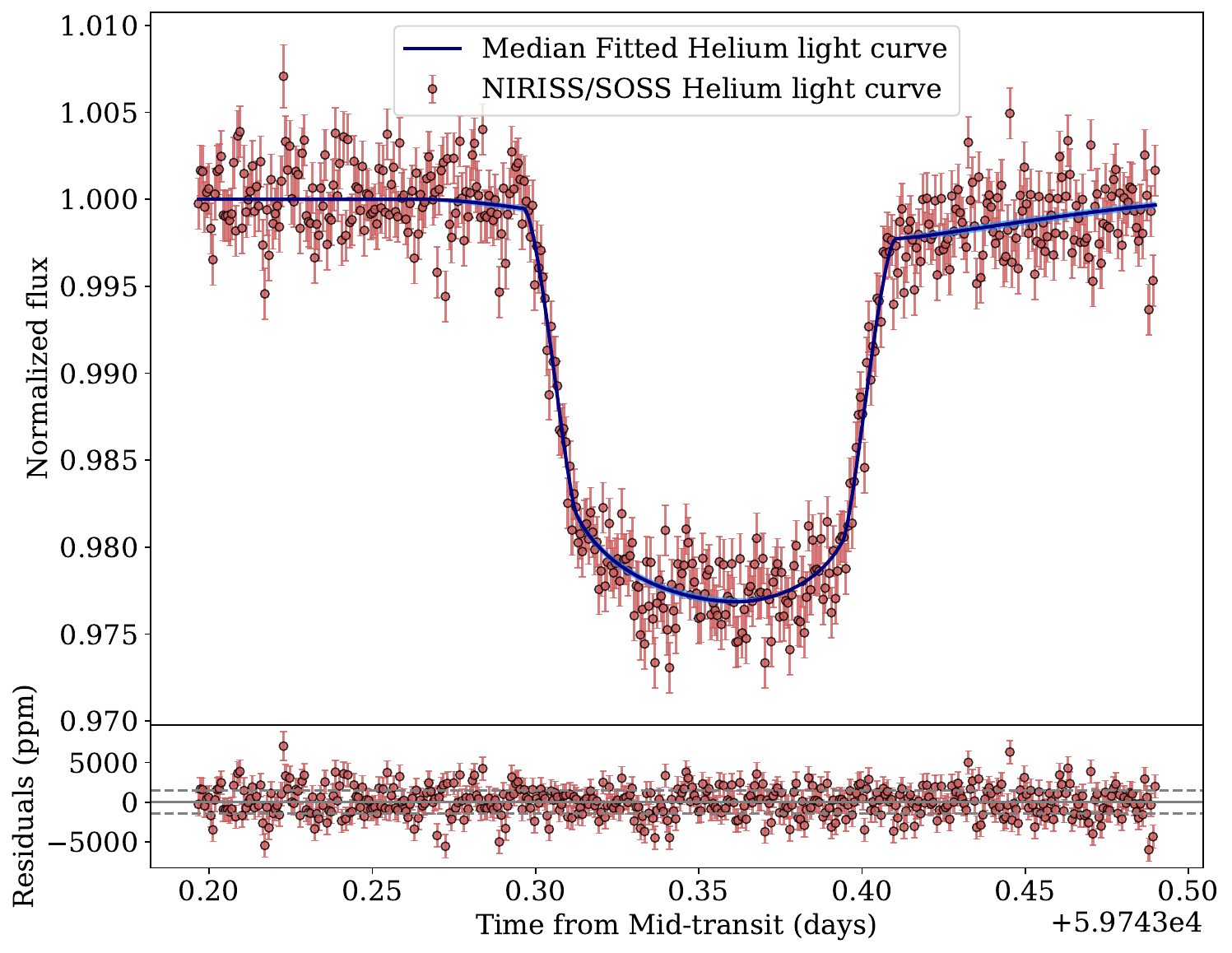} 
    \includegraphics[width=0.475\linewidth]{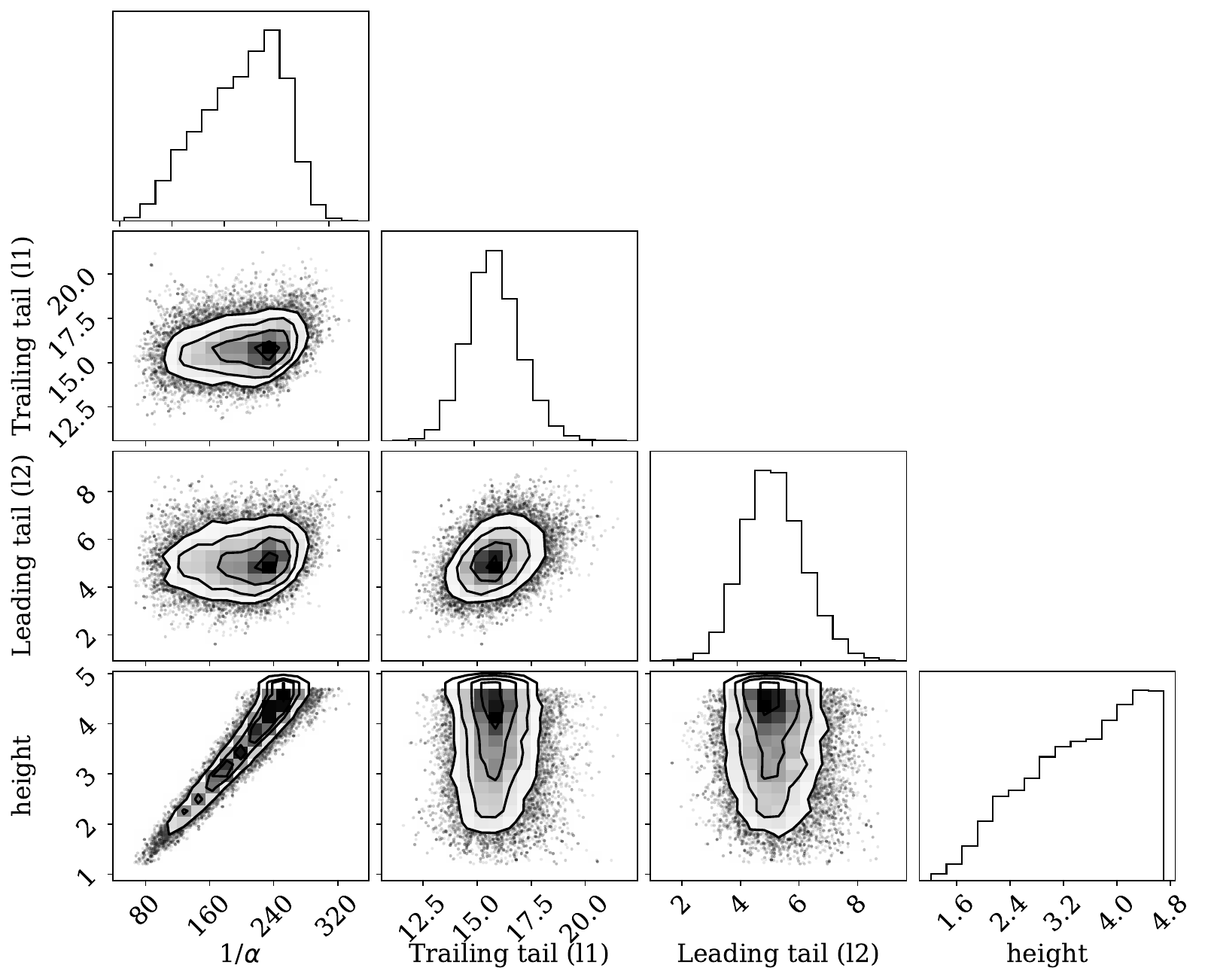}
    \includegraphics[width=1.\linewidth]{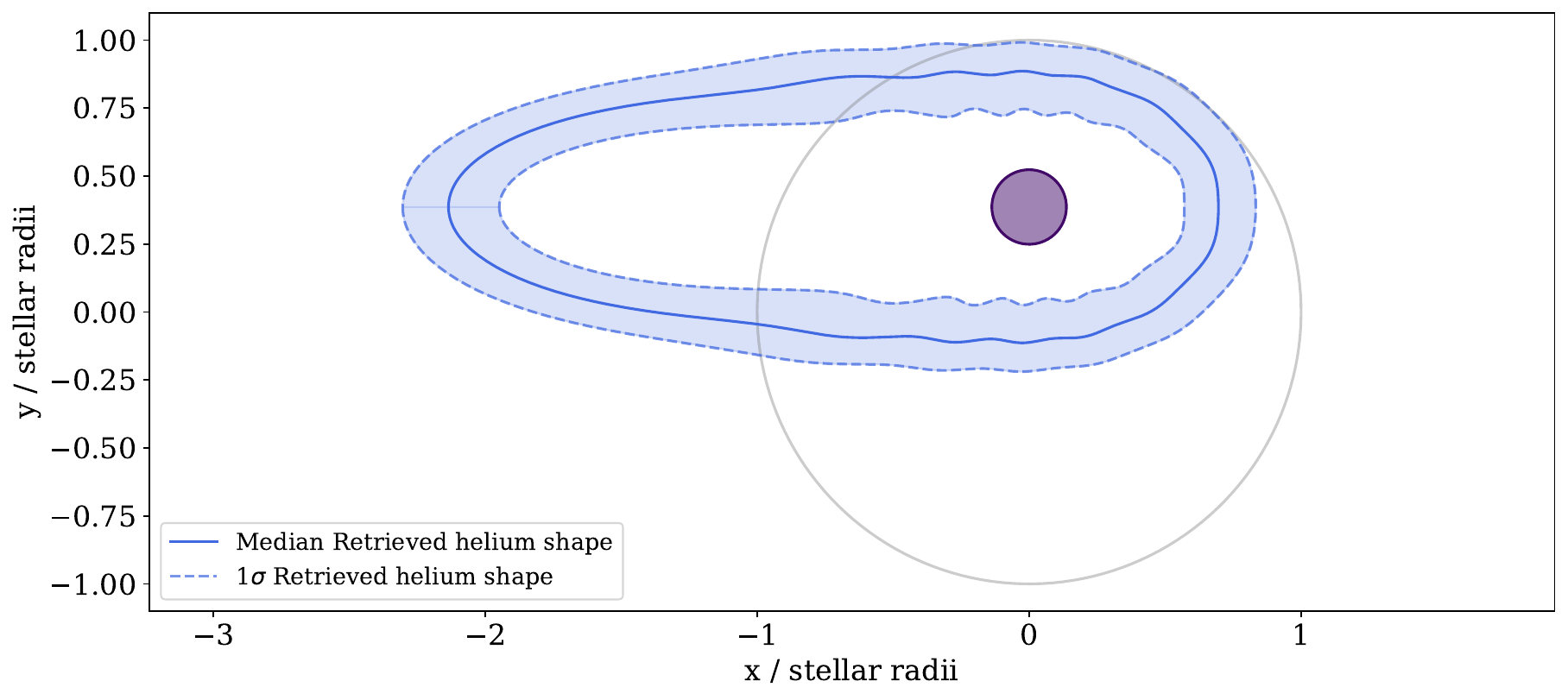}
    \caption{Main results of HAT-P-18b uniform helium envelope fit. The top left panel shows the NIRISS/SOSS light curve in red, corresponding to the 10\AA~bin centered around the He triplet, together with the fitted model light curve in blue. The top right panel shows the posteriors of the four fitted parameters ($l_1$,  $l_2$, $h$, $\alpha$). The bottom panel show the median retrieved envelope in blue, with the shaded region showing the $1\sigma$ uncertainty region. The gray line represents the star and the purple circle illustrates the planet.}
    \label{fig:h18}
\end{figure*}

We apply our method to the helium escape observation of HAT-P-18b reported by \cite{Fu2022} and \cite{fournier2024} using the JWST/NIRISS SOSS observations of this planet obtained as part of the Early Release Observations (ERO) program 2734 (PI: Pontoppidan). The observations consisted of a continuous 7.16 hours time series -- 469 integrations, 54.94 seconds each -- of this target, that contain  the 2.71 hour planet transit, plus 4.45 hours of baseline equally split before and after transit. We reduced the observations using the \texttt{transitspectroscopy} pipeline \citep{espinoza_nestor_2022_6960924}.  
Starting from the \texttt{rateints.fits} files produced by the \texttt{jwst} pipeline, we first remove the background flux using the model background image created during JWST's commissioning. We scale the model image using the pixels located in a region of the image uncontaminated by the target's trace, and then subtract the scaled model from every integration. We correct the 1/f noise at the integration level by first removing the median frame from each integration, and subsequently calculating and subtracting to each column the median of the pixels surrounding the first order trace. Finally, we extract the spectra using a box extraction with a 15 pixels half-width aperture, and correct for cosmic rays and bad pixels on the extracted spectra by flagging and replacing 5-sigma outliers in time.
We focus our analysis on the first order spectrum ($0.8$ to $1.7\micron$) as it contains the 10830\AA~metastable helium triple line. Figure \ref{fig:h18_he_wlc} in the appendix \ref{sec:app_h18} shows the first order broadband (white) light curve in dark blue, compared to the light curve given by the $\sim$10\AA~bin centered around the He triplet in red. Compared to the broadband light curve, the He triplet light curve shows both a deeper transit depth and significant asymmetry between ingress and egress, which suggests the presence of a prominent He tail. We fit this tail following the same procedure described above for a single uniform envelope model. 

\subsection{Uniform envelope fit}
We fit the helium light curve of HAT-P-18b assuming the transit of the planet surrounded by an envelope of escaping gas parametrized by the trailing and leading tail sizes, $l_1$ and $l_2$, the height parameter, $h$,  and an average envelope opacity, $\alpha$. We fix all other system parameters to the values reported in \cite{Fu2022}, which are given in Table \ref{tab:sysparams}. We compute the limb-darkening coefficients using the \texttt{ExoTiC-LD} package \citep{Grant2024}, using the quadratic law, the 3D models from \cite{Magic2015} and the stellar parameters reported in \cite{hartman2011}. We perform the fits using the Markov Chain Monte Carlo (MCMC) sampler \texttt{emcee} \citep{emcee}, running 20 walkers, each with 400 steps as burn-in and 2000 steps for the production run. For all four parameters, we adopt uniform priors. 

The top right panel in Figure \ref{fig:h18} shows the posterior distributions of the four fitted parameters, while the lower panel illustrates the median retrieved envelope (blue solid line), together with the corresponding range of uncertainty (blue shaded region). The residuals, plotted in the upper left panel of Figure \ref{fig:h18}, show no clear signs of remaining systematics, evidencing that our envelope model is capable of modeling the asymmetry in the data. As a result, we are able to infer a trailing tail size of $l_1 = 15.79^{+1.14}_{-1.05} R_p$ and a leading tail size of $l_2 = 5.11^{+0.99}_{-0.89} R_p$. We observe that the post transit absorption of HAT-P-18b never settles, but instead appears to rise throughout the entire post-transit baseline (Figure \ref{fig:h18}). In this regard, we note that the inferred tail sizes could potentially be higher, as more pre- or post-transit absorption could have been missed by the insufficient baseline captured with these observations. On the other hand, although there is no apparent pre-transit absorption (see Figure \ref{fig:h18_he_wlc}), we are still able to infer a significant leading tail in HAT-P-18b's helium envelope. This indicates that the general shape of the light curve, including the in-transit features, is a contributing factor in the inference of the envelope morphology. We further discuss this in Appendix \ref{sec:app_h18}. Finally, as in the tests in Section \ref{sec:injection}, we observe a strong degeneracy between $h$ and $\alpha$, which hinders our ability to accurately measure these two parameters. Nonetheless, we are still able to constrain the envelope total effective area $A_{\text{eff}} = 0.562^{+0.036}_{-0.035} R_{p}^2$, as well as the tail ratio $l_1/l_2 = 3.08^{+0.60}_{-0.45}$ . The fitted parameters are summarized in Table \ref{tab:results}. 

We have also tried performing a multi-layer fit to the HAT-P-18b data, using $M = 2$ layers. While the inner layer is in good agreement with the results from the uniform envelope case, the outer layer is highly unconstrained. Furthermore, the multiple layer fit has a significantly higher AIC compared to the single layer fit. This is consistent with the results from the injection recovery tests, which showed that there is no information to gain from adding multiple layers to the fit given the precision of the observations ($\sim 2000$ ppm). Finally, to evaluate the proposed methodology on ground-based data, we have also analyzed the Keck/NIRSPEC observations of WASP-107b in Appendix \ref{app_w107}.

\section{Conclusions} \label{sec:conclusions}

In this work we have presented a framework to fit the assymetric light curves resulting from the transit of a planet surrounded by an extended tail of escaping gas, in order to infer the size and geometry of the latter. We have defined the necessary formulation to parametrize the envelope of escaping helium around a planet, and proposed a methodology to calculate the corresponding transit light curves using the publicly available software \texttt{Harmonica} \citep{grant2022transmission}. We have performed injection-recovery tests on synthetic data to validate the robustness of the method presented. We have shown that, with light curve precisions similar to those achieved by NIRISS/SOSS, it is possible to constrain the size of the leading and trailing tails with an error of $\sim 1 R_p$. We have applied the proposed methodology to the NIRISS/SOSS observations of HAT-P-18b, which we reduced using the \texttt{transitspectroscopy} pipeline \citep{espinoza_nestor_2022_6960924}. As a result, we conclude that the planet must posses a trailing tail of $15.79^{+1.14}_{-1.05} R_p$ and a leading tail of $5.11^{+0.99}_{-0.89} R_p$, as well as an effective area of $A_{\text{eff}} = 0.562^{+0.036}_{-0.035} R_{p}^2$. Additionally, we have analyzed the Keck/NIRSPEC observations of WASP-107b, demonstrating that our method can also be applied to ground-based observations.

\section{Discussion and Future Work} \label{sec:futurework}
We summarize here some of the future work that we aim to carry out as an improvement or application of the ideas presented in this letter:
\begin{itemize}
    \item The representation of large tails of escaping gas poses some challenges to the circular harmonic parametrization used in \texttt{Harmonica}, specially for very slim shapes for which high number of harmonics are needed. One solution would consist in adding a elliptical harmonic basis to \texttt{Harmonica}'s framework. This would not only allow us to represent the elliptical envelope with a much lower number of harmonics, but would also open the door to the consideration of more complex shapes thanks to \texttt{Harmonica}'s ability to compute the transit light curve of any shape. Alternatively, we could consider developing a fixed shape code, similar to \texttt{batman} \citep{kreidberg2015} or \texttt{catwoman} \citep{espinoza2021}, for the two-ellipse shape presented in this study. While developing these solutions is out of the scope of this letter, we consider these as part of the future work we aim to carry out in the context of this study.

    \item In general, the ability to constrain the tail parameters will depend on several variables such as the actual shape and size of the helium envelope, the underlying system parameters, and multiple instrumental factors (e.g.,  precision and resolution of the observation). For instance, deriving the envelope properties might be difficult for planets with a high impact parameter $b$, for which the transit chord is short and a significant portion of the envelope might not transit the star. The opacity of the envelope will also determine the effectiveness of our method. In this regard, very opaque envelopes will be responsible for larger helium signals, and will therefore be easier to characterize. In future studies we plan to extensively investigate how the system parameters determine our ability to constrain the envelope morphology.

    \item We note that our model reconstructs a sky-projected 2D map of the surface opacity, therefore compressing information along the line of sight to the star. This means that there are multiple 3D configurations of material along the line of sight that could satisfy the same sky-projected 2D distribution. While it is true that the shape of the outflow in the direction normal to the sky plane could in some cases affect the transit morphology, we generally expect the effects to be small and difficult to probe given the precision of the observations and the typical geometric configuration of the transits. In some cases in which the absorption structure along the line of sight is large enough (e.g Roche-lobe overflow), the normal flow might be revealed during ingress or egress, hence allowing our model to extract some information on the 3D distribution of the escaping envelope. We plan to investigate this in future studies.
    
    \item Currently there are more than $\sim$50 planets with NIRISS/SOSS observations covering the helium triplet line at 10830\AA, including planets with known strong helium escape signals such as WASP-107b \citep{spake2018} or WASP-69b \citep{nortmann2018, tyler2024}. The approved JWST archival program entitled \textit{The Unintentional NIRISS Escape Survey} (TUNES) (AR 5916, PI: Vissapragada) is uniformly analyzing all the publicly available NIRISS/SOSS observations in order to produce the largest helium escape survey to date. As part of TUNES's analysis plan, we will apply the methodology presented in this paper to a large sample of exoplanets in different star-planet environments. As a result, we aim to conduct a population study in which we compare the different helium envelope morphologies to the corresponding stellar and planetary parameters.
    
    \item The relatively low resolution of NIRISS/SOSS (R $\sim$ 700) significantly dilutes the helium signal and therefore limits the conclusions from our morphological study. Alternatively, JWST's NIRSpec G140H grism is able to sample the helium triplet line at 10830\r{A} with 4 times the resolving power (R $\sim$ 2700), while achieving precisions comparable to those obtained with NIRISS/SOSS. Observations with NIRSpec's G140H grism would therefore yield significantly stronger helium signals, which would in turn facilitate and improve the morphological study of the corresponding escaping envelope \citep{dossantos2023}. 
    
    \item Ground-based observations have been the workhorse for detecting and studying the presence of escaping helium in exoplanet atmospheres. Despite the difficulties introduced by telluric contamination and atmospheric variability, the higher resolution achievable with ground-based spectrographs opens the door to the detection and characterization of significantly stronger helium signals. In Appendix \ref{app_w107} we have demonstrated that our methodology can be used to fit ground-based data too. In this regard, we also plan to analyze the helium light curves observed with ground-based high-resolution spectrographs such as KECK/NIRSpec, CARMENES or HET/HPF.
\end{itemize}

\appendix 
\restartappendixnumbering 

\section{Examples of escaping tails and morphologies} \label{sec:app_ex}

\begin{figure*}
    \centering
    \includegraphics[width=0.49\linewidth]{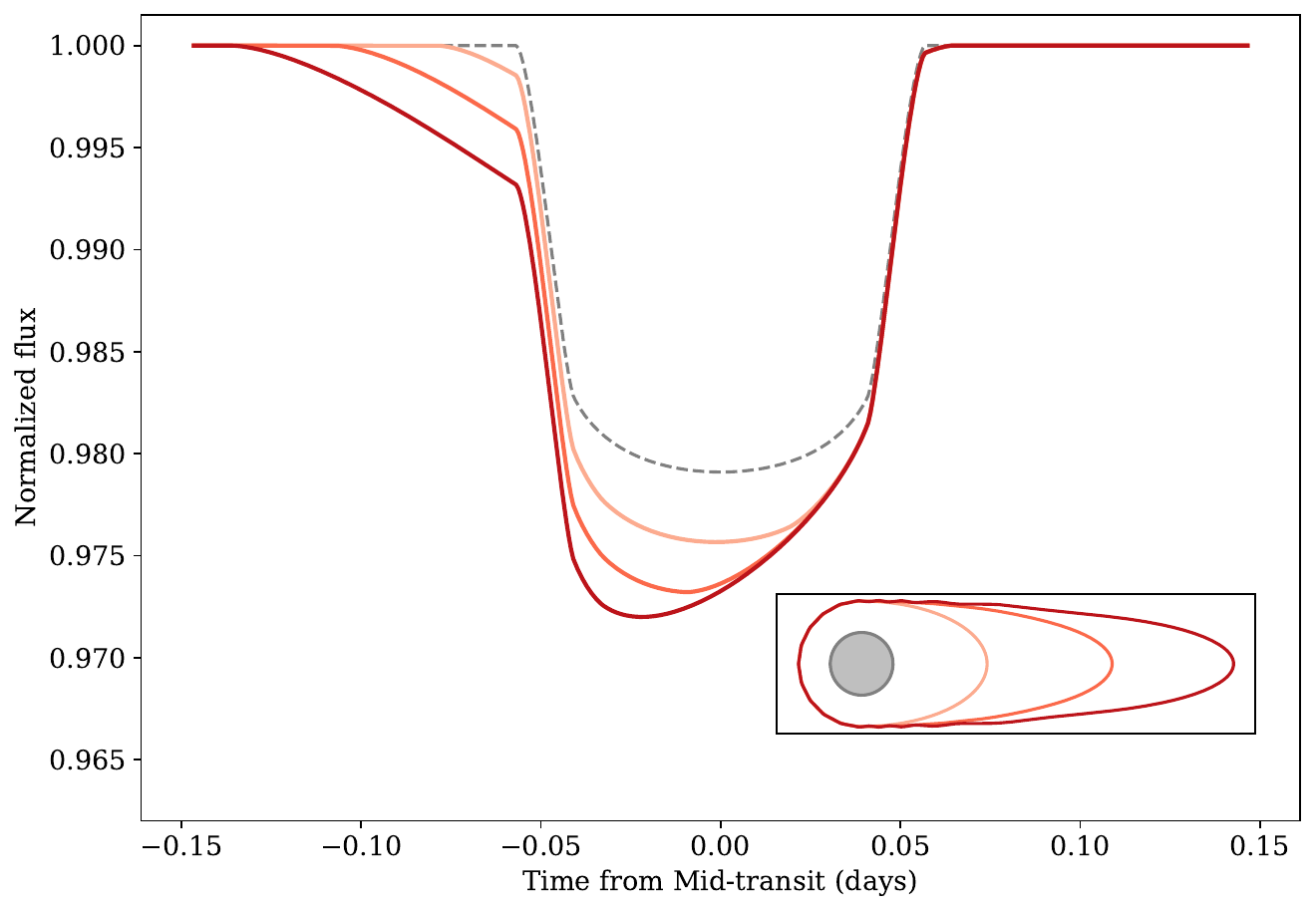} 
    \includegraphics[width=0.49\linewidth]{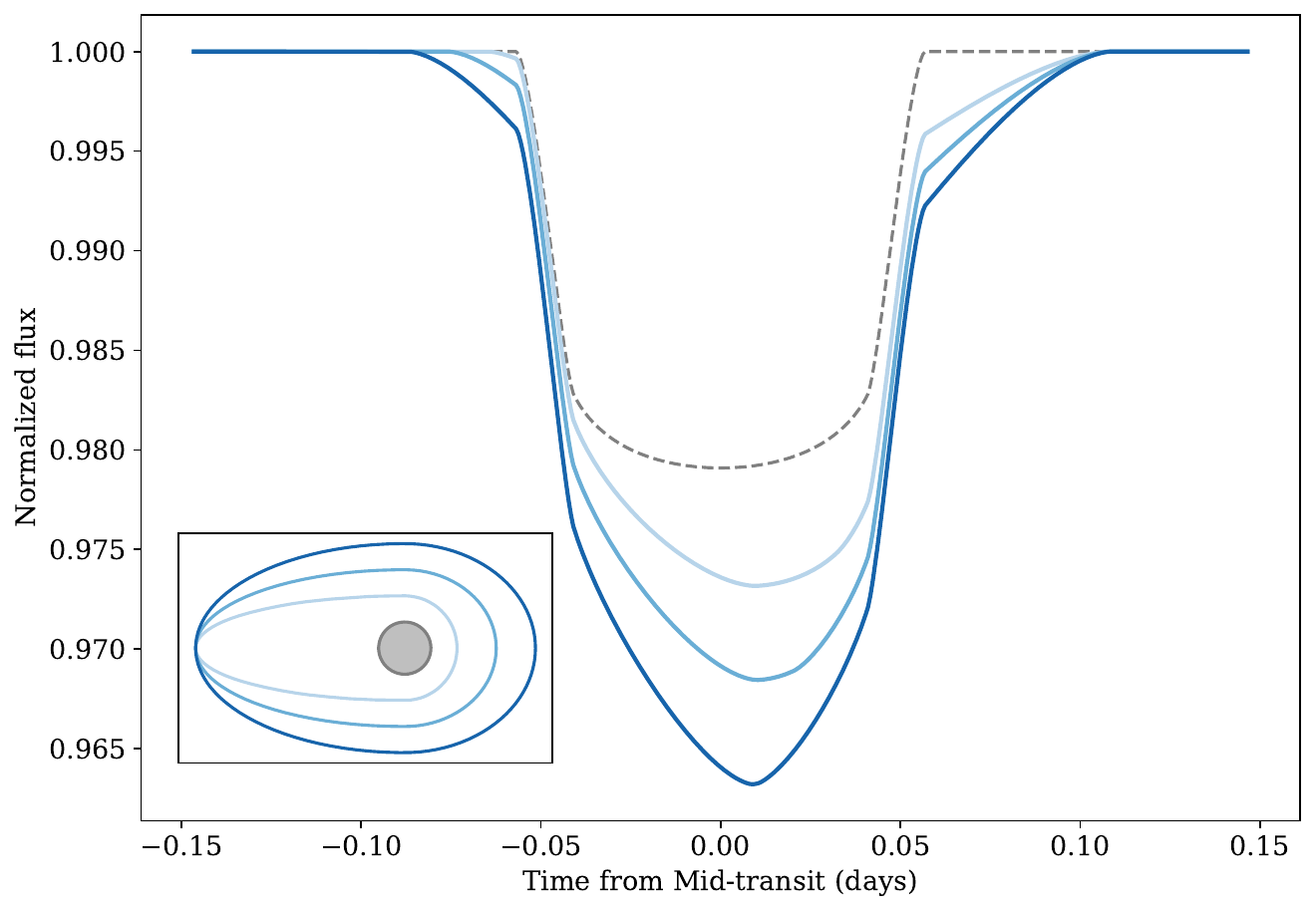}
    \caption{Transit light curves corresponding to different envelope morphologies.  The left panel shows several envelopes with a prominent increasing leading tail. The height $h$ and trailing tail $l_1$ are kept constant. The right panel, on the other hand, shows envelopes with a constant prominent trailing tail, but with an increasing height and leading tail. In all cases, the opacity is left constant.}
    \label{fig:examples}
\end{figure*}

\begin{figure*}
    \centering
    \includegraphics[width=0.49\linewidth]{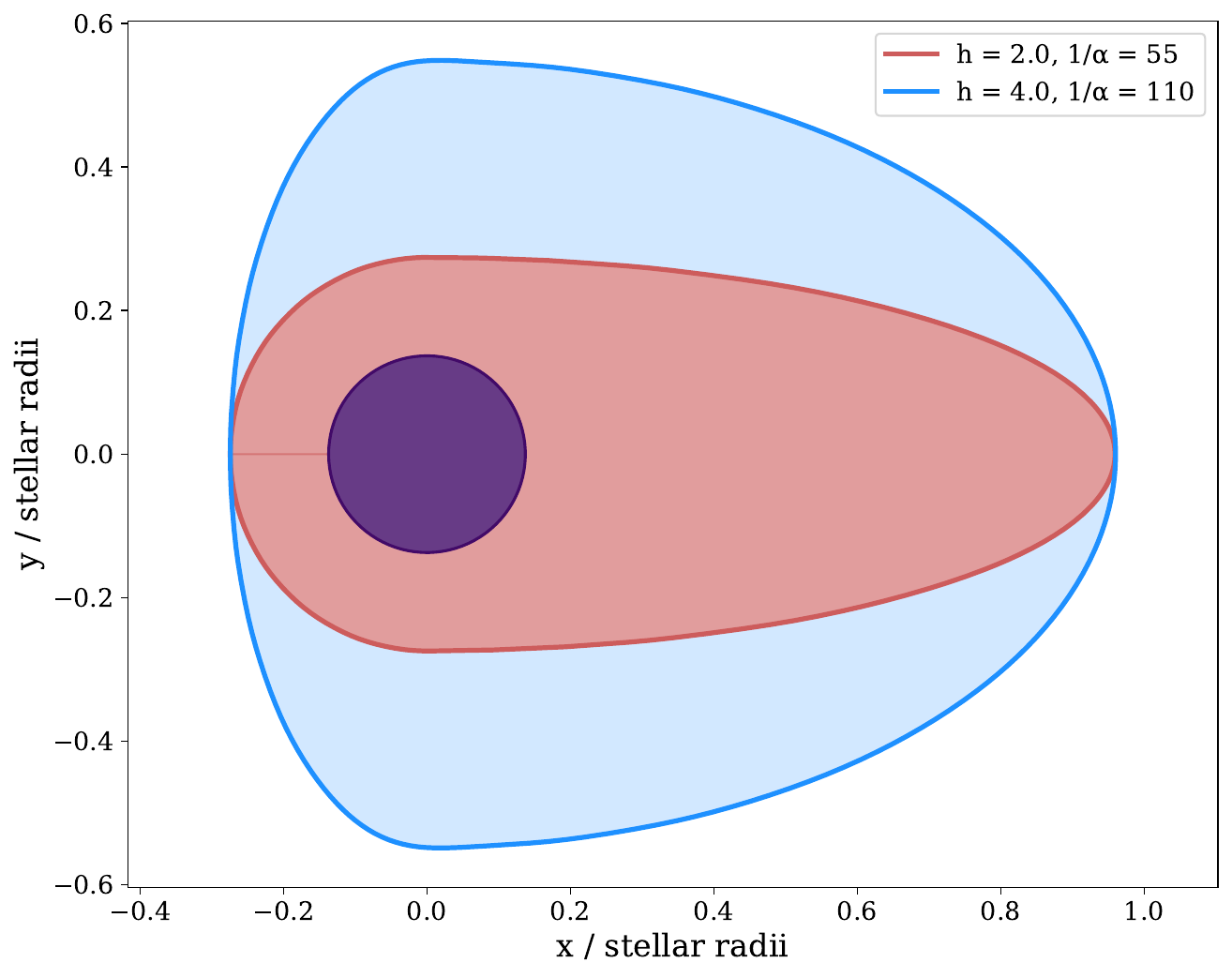} 
    \includegraphics[width=0.49\linewidth]{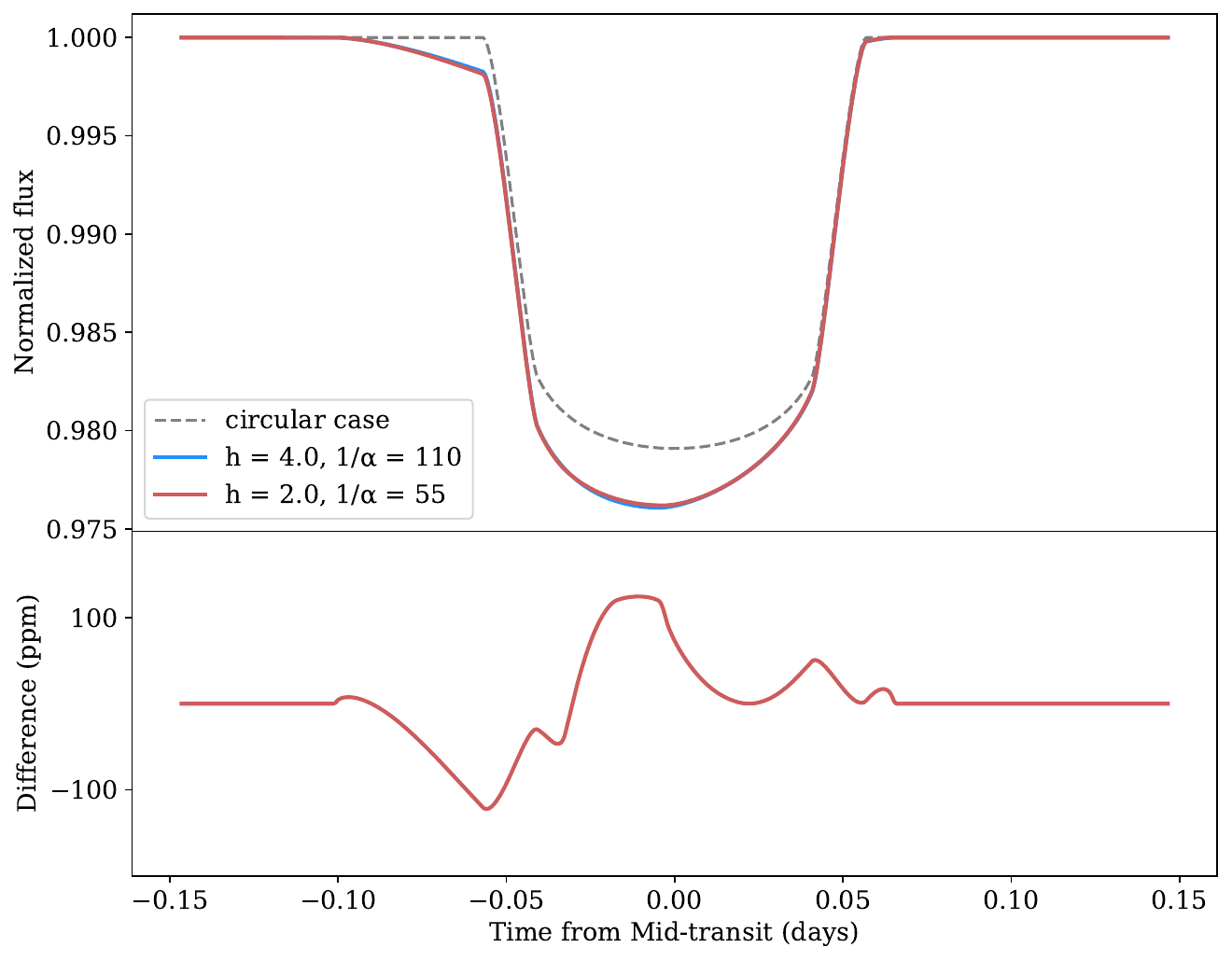}
    \caption{Example of the degeneracy between the envelope height $h$ and the opacity $\alpha$. The red shaded region illustrates a smaller more opaque envelope, while the blue shaded region represents a larger more transparent envelope. The right panel shows the corresponding light curves, with the difference plotted in the lower panel.}
    \label{fig:degeneracy}
\end{figure*}

In Figure \ref{fig:examples} we show the transit light curves corresponding to different envelope morphologies. The left panel shows several envelopes with a prominent increasing leading tail. The height $h$ and trailing tail $l_1$ are left constant. The right panel, on the other hand, shows envelopes with a constant prominent trailing tail, but with an increasing height and leading tail. In all cases, the opacity is fixed to the same value.
Figure \ref{fig:degeneracy} shows an example of the degeneracy between the envelope height and the opacity. The left panel shows two envelopes with the same effective area, but with different shapes and opacities. In particular, the red shaded region illustrates a smaller more opaque envelope, while the blue shaded region represents a larger more transparent envelope. As we can observe, the corresponding transit light curves (shown in the right panel) are very similar, with differences only of the order of 100 ppm.

\section{Injection recovery test figures and tables}\label{sec:app_inj}
\begin{figure*}
    \centering
     \includegraphics[width=0.5\linewidth]{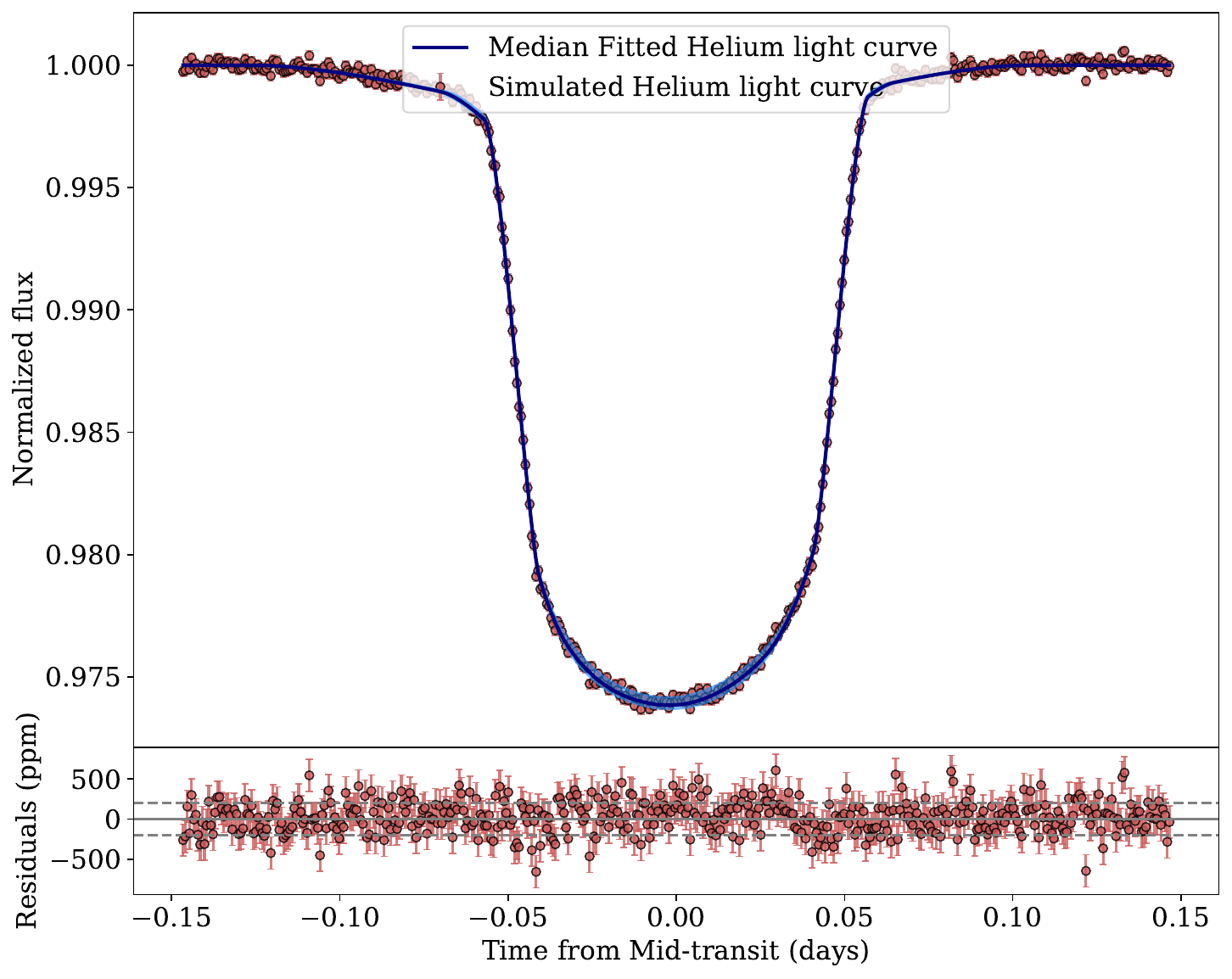} \\
    \includegraphics[width=0.95\linewidth]{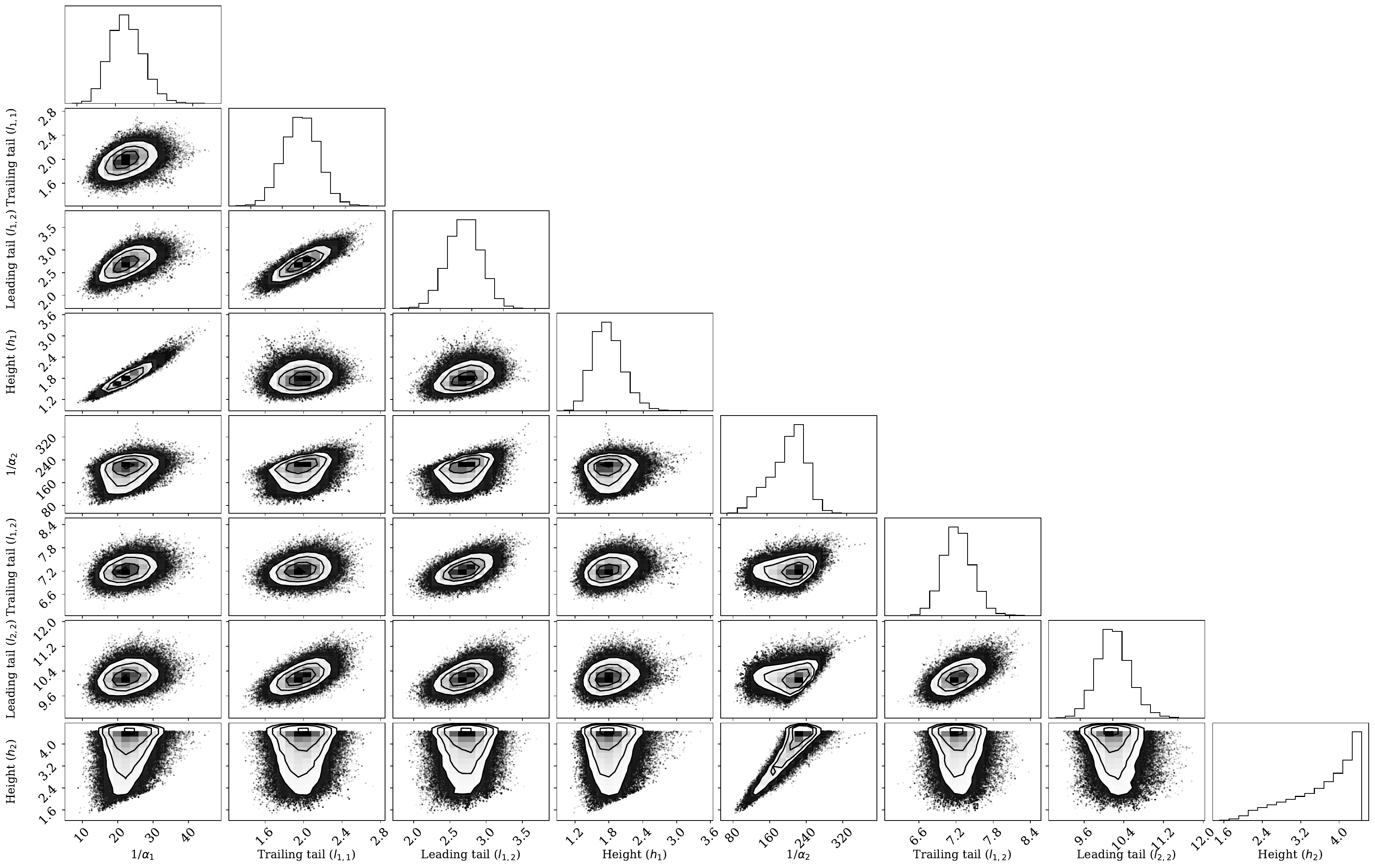} 
    \caption{Main results of the injection-recovery test for the 2 layer, 200 ppm precision case. The top panel shows the injected light curve in red together with the fitted model light curve in blue, while the bottom panel shows the posterior distribution of the fitted parameters.}
    \label{fig:multi_layer_results}
\end{figure*}

In this appendix we include additional figures and tables to complement the injection recovery test presented in Section \ref{sec:injection}. Figure \ref{fig:multi_layer_results} shows the main results of the multi-layer envelope fit corresponding to $M=2$ layers and the $200$ppm precision case. The bottom panel shows the posterior distribution of the fitted parameters. The top panel presents the simulated light curve in red, and the fitted light curve in blue. Finally, Table \ref{tab:results_layers} summarizes the results from the multi-layer analysis presented in Section \ref{sec:multilayer}. For each precision and number of layers $M$ we report the derived values of the effective area and the Akaike Information Criterion (AIC).

\begin{deluxetable*}{l|c|cc|cccc}
\tablecaption{Results from the injection-recovery test for the helium envelope with a variable opacity. For each precision and number of layers tested, we report the corresponding effective area and AIC value.}
\label{tab:results_layers}
\tablehead{\colhead{}  & \colhead{Injected} & \colhead{1 layer} & \colhead{2 layers} & \colhead{1 layer} & \colhead{2 layers} & \colhead{3 layers} & \colhead{4 layers}\\
\colhead{} & \colhead{} & \colhead{(2000 ppm)} & \colhead{(2000 ppm)} & \colhead{(200 ppm)} & \colhead{(200 ppm)} & \colhead{(200 ppm)} & \colhead{(200 ppm)}}
\startdata
$A_{\text{eff}}$ &  0.889 & 0.894$^{+0.046}_{-0.044}$ & 0.900$^{+0.045}_{-0.045}$ & 0.838$^{+0.006}_{-0.005}$ & 0.881$^{+0.011}_{-0.014}$ & 0.874$^{+0.012}_{-0.010}$ & 0.872$^{+0.011}_{-0.010}$ \\
AIC  & - & -2221  & -2214  & -3039 & -3260 & -3256 & -3254 
\enddata
\end{deluxetable*}

\onecolumngrid
\section{HAT-P-18b FIGURES AND TABLES}\label{sec:app_h18}
\begin{figure*}
    \centering
    \includegraphics[width=0.58\linewidth]{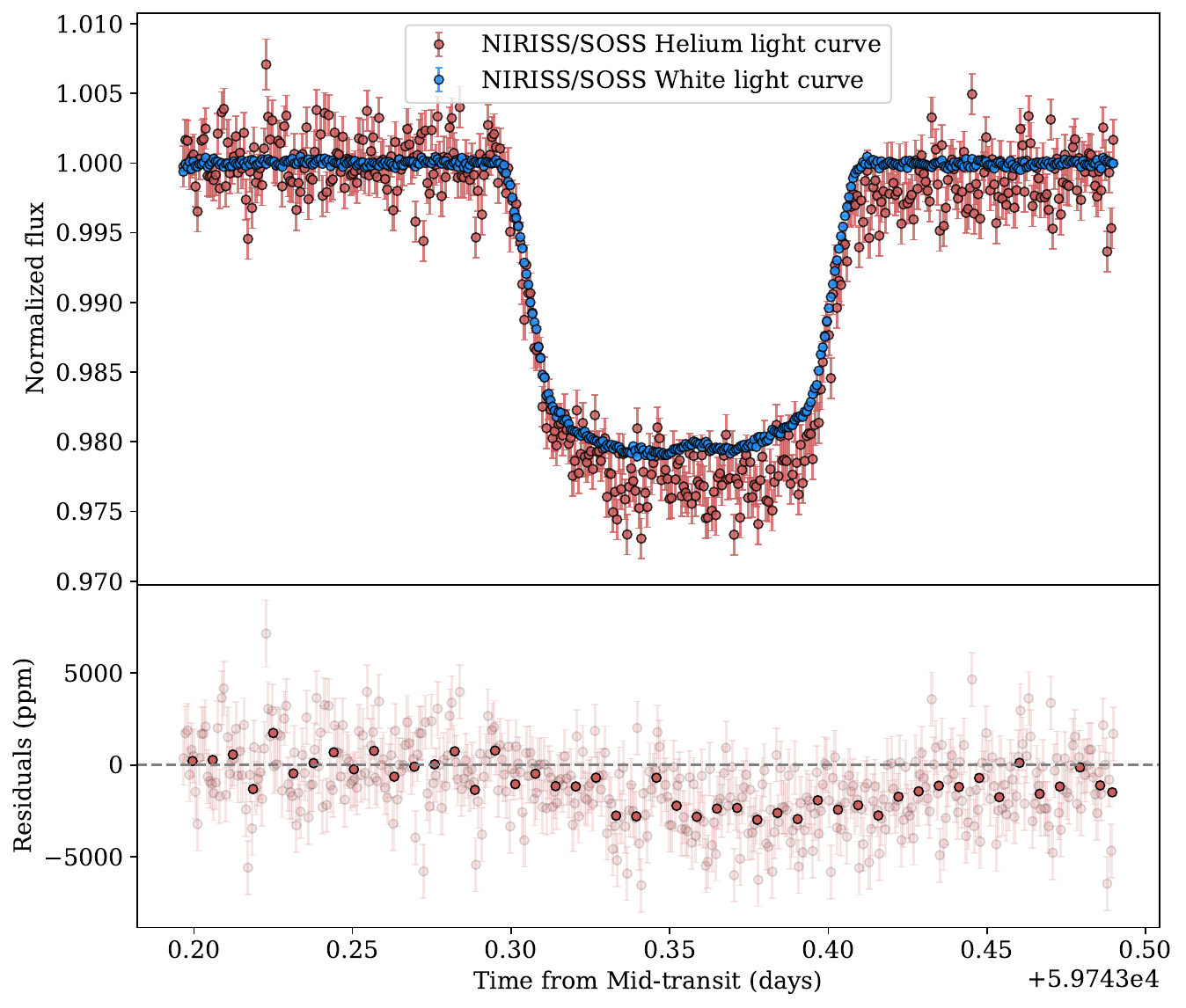}
    \caption{HAT-P-18b extracted light curves. The top panel shows the 10\AA~bin light curve centered around the He triplet in red, together with the white light curve in blue. The lower panel shows the binned residuals, illustrating both the larger depth and asymmetry present in the helium light curve. }
    \label{fig:h18_he_wlc}
\end{figure*}

\begin{figure*}
    \centering
    \includegraphics[width=0.51\linewidth]{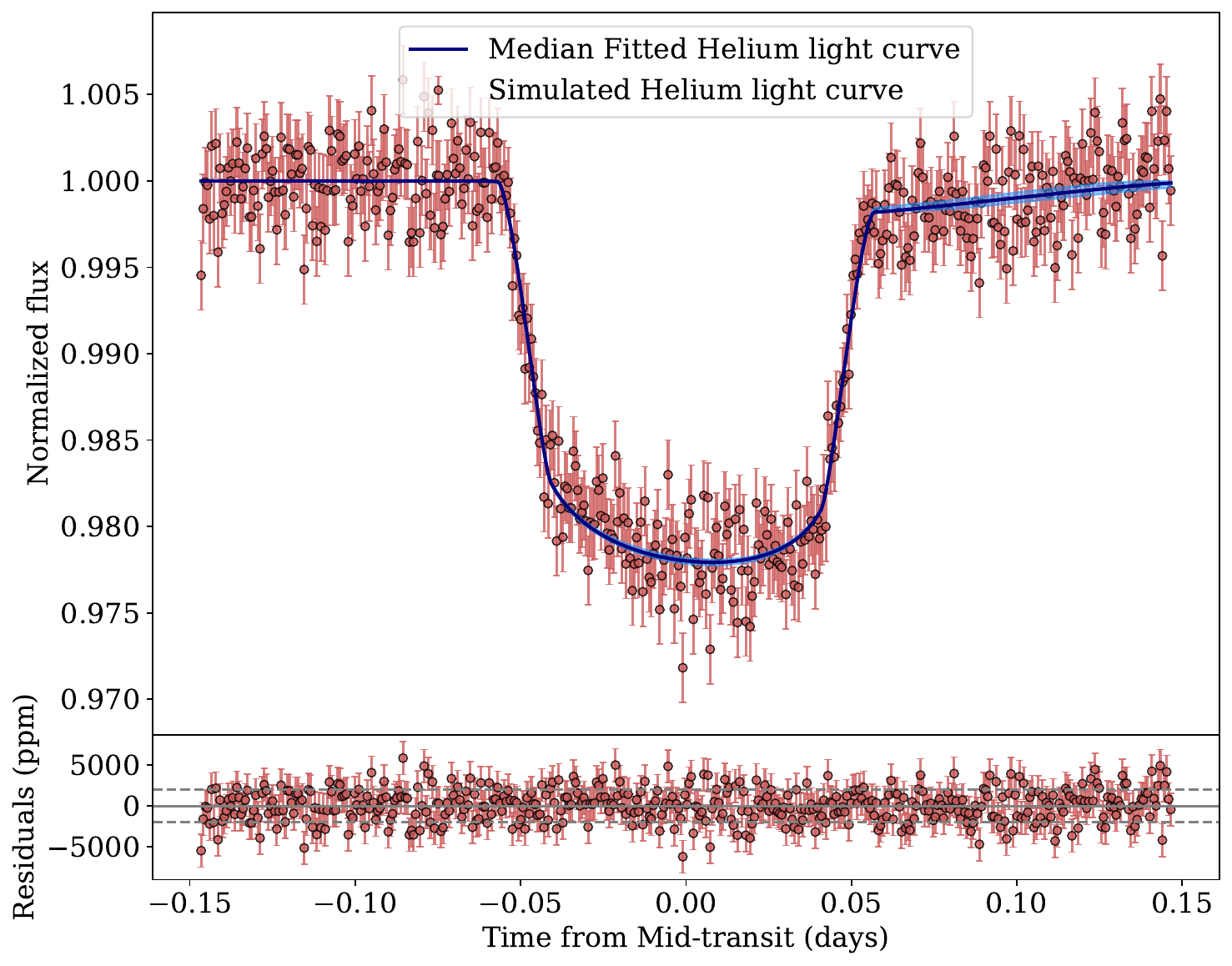} 
    \includegraphics[width=0.48\linewidth]{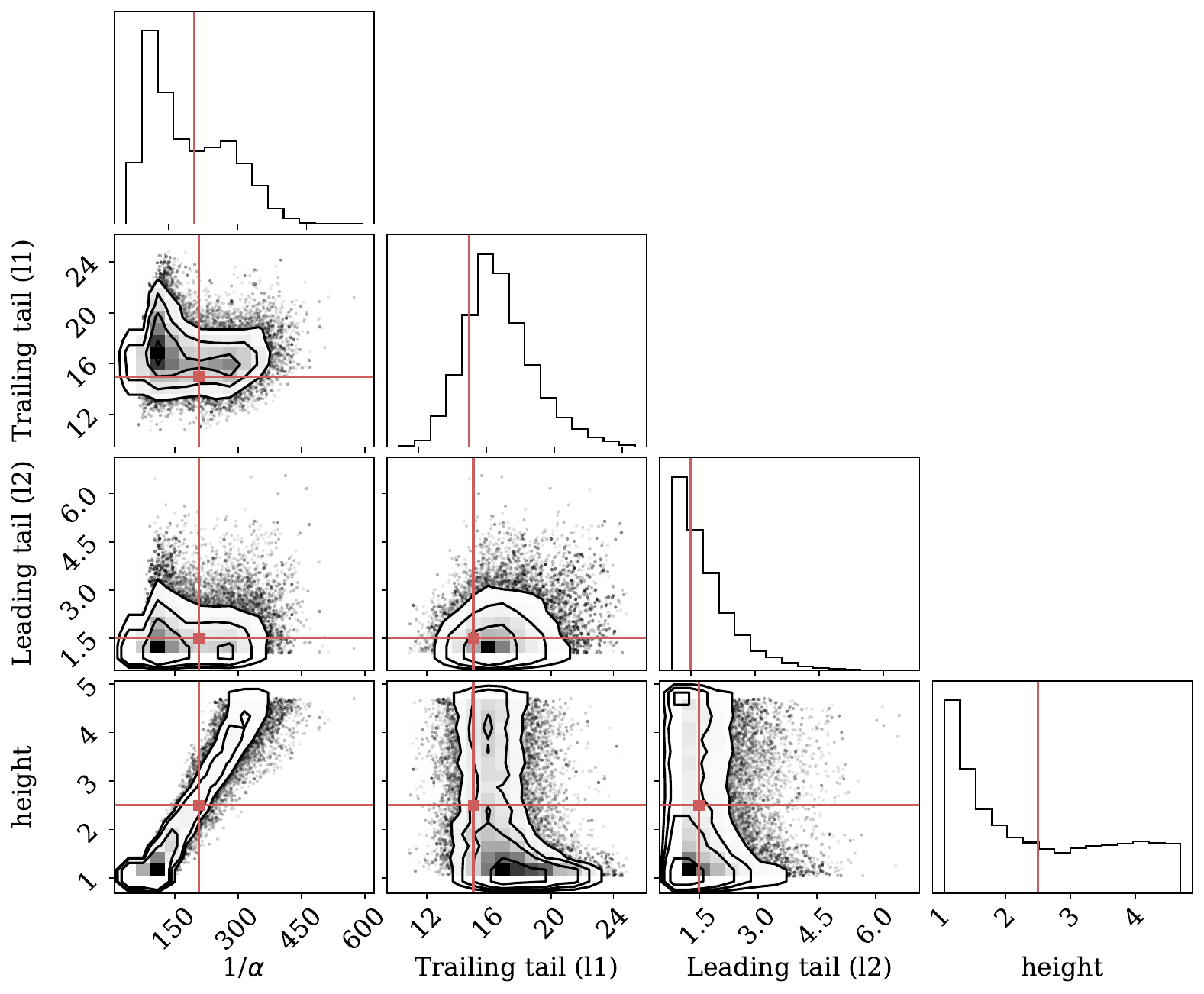}
    \caption{Results from fitting a simulated HAT-P-18b light curve, assuming the envelope has no leading tail. The left panel shows the simulated light curve in red, together with the fitted model in blue. The right panel shows the posteriors of the four fitted parameters, with the injected values shown in red.}
    \label{fig:h18_sim}
\end{figure*}

In this appendix we include additional figures and tables corresponding to the analysis of the HAT-P-18b observations presented in Section \ref{sec:application}. The top panel in Figure \ref{fig:h18_he_wlc} compares HAT-P-18b's broadband light curve in dark blue, computed by integrating all wavelengths from 0.8 to 2.8 $\micron$, with the helium light curve in red. When compared to the white light curve, the helium light curve exhibits not only a larger transit depth, but also a clear asymmetry mainly driven by post transit absorption. This is further evidenced in the bottom panel, where we show the residuals resulting from taking the difference between both light curves. Figure \ref{fig:h18_he_wlc} shows no apparent pre-transit absorption, which would suggest no leading tail in HAT-P-18b's helium envelope. However, the analysis performed in Section \ref{sec:application} yields a significant leading tail  of $l_2 = 5.11 ^{+1.14}_{-1.05} R_p$. To further investigate this, we perform an additional injection-recovery test where we simulate a light curve similar to that of HAT-P-18b with parameters $l_1 = 15 R_p$, $h = 2.5 R_p$ and $1/\alpha = 207$, but with a significantly smaller leading tail $l_2 = 1.5 R_p$. Figure \ref{fig:h18_sim} shows the simulated light curve and the fitted model in the left panel, together with the corresponding posteriors and injected values in the right panel. We retrieve a  leading tail of $1.6^{+0.8}_{-0.4} R_p$, which is in very good agreement with the injected value. Essentially, this demonstrates that the tail sizes inferred for the real HAT-P-18b observations are indeed statistically robust. Even if there is no evident pre or post-transit absorption, our methodology is able to accurately extract the information embedded in the general shape of the transit light curve. Table \ref{tab:sysparams} summarizes HAT-P-18b's system parameters, extracted from \cite{Fu2022}.

\begin{deluxetable*}{lll}
\tablecaption{HAT-P-18b System Parameters extracted from \cite{Fu2022}. The period is extracted from Kokori et al. 2023}
\label{tab:sysparams}
\tablehead{\colhead{Parameter} & \colhead{Value} & \colhead{Description} }
\startdata
$P$ & 5.50802941 & Orbital Period (days)\\
$a/R_{*}$ & 16.52 & Scaled Semi-major axis \\
$i$ & 88.66 & Inclination (deg) \\
$e$ & 0 & Eccentricity \\
$t_0$ & 2459743.853395 & Time of Mid-transit (days, MJD) \\
$R_p$ & 0.137113\tablenotemark{a} & Planetary Radius \\
\enddata
\tablenotetext{a}{The reported value corresponds to a planetary radius measured outside the helium band as stated in \citep{Fu2022}}
\end{deluxetable*}

\onecolumngrid
\section{Application to ground-based data}\label{app_w107}

\begin{figure*}
    \centering
    \includegraphics[width=0.49\linewidth]{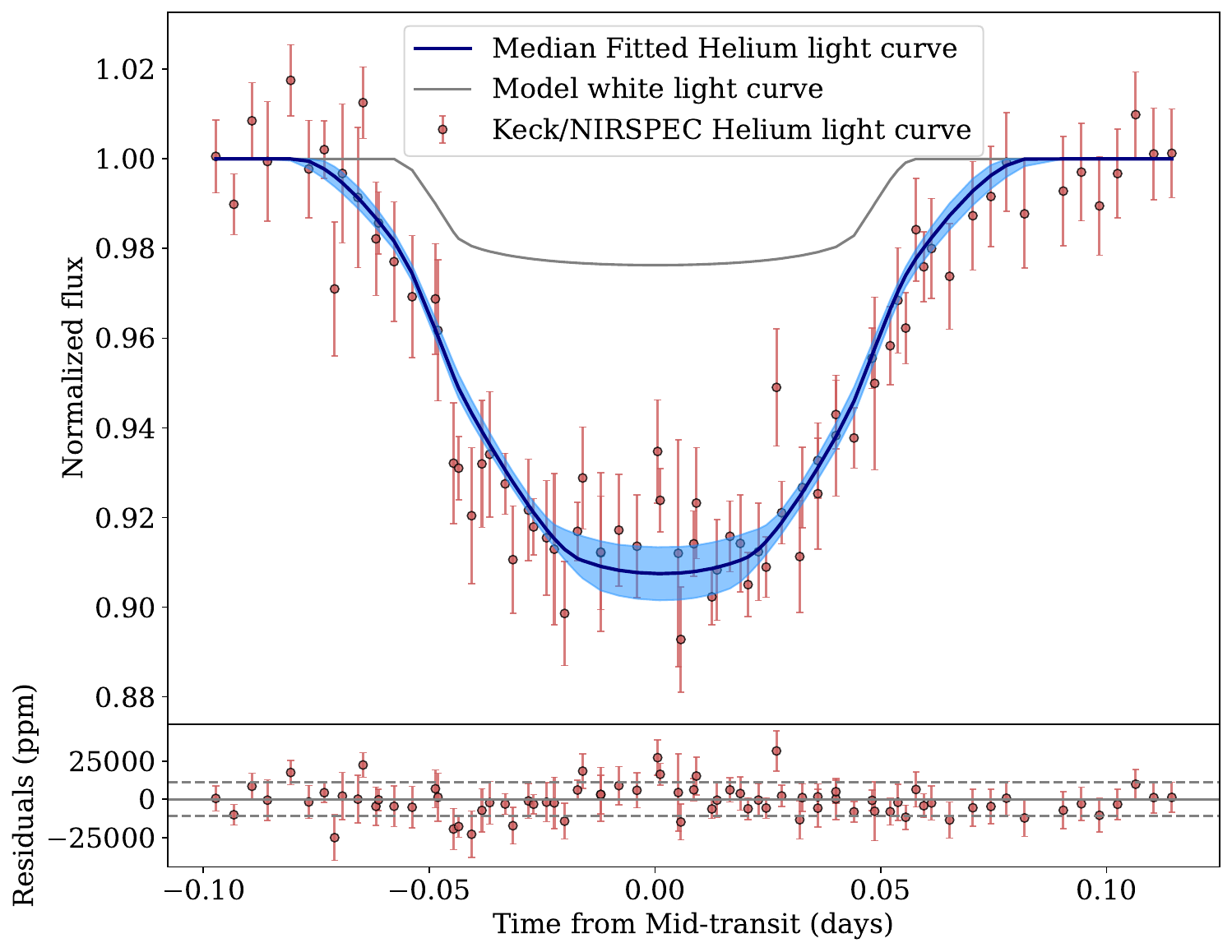} 
    \includegraphics[width=0.49\linewidth]{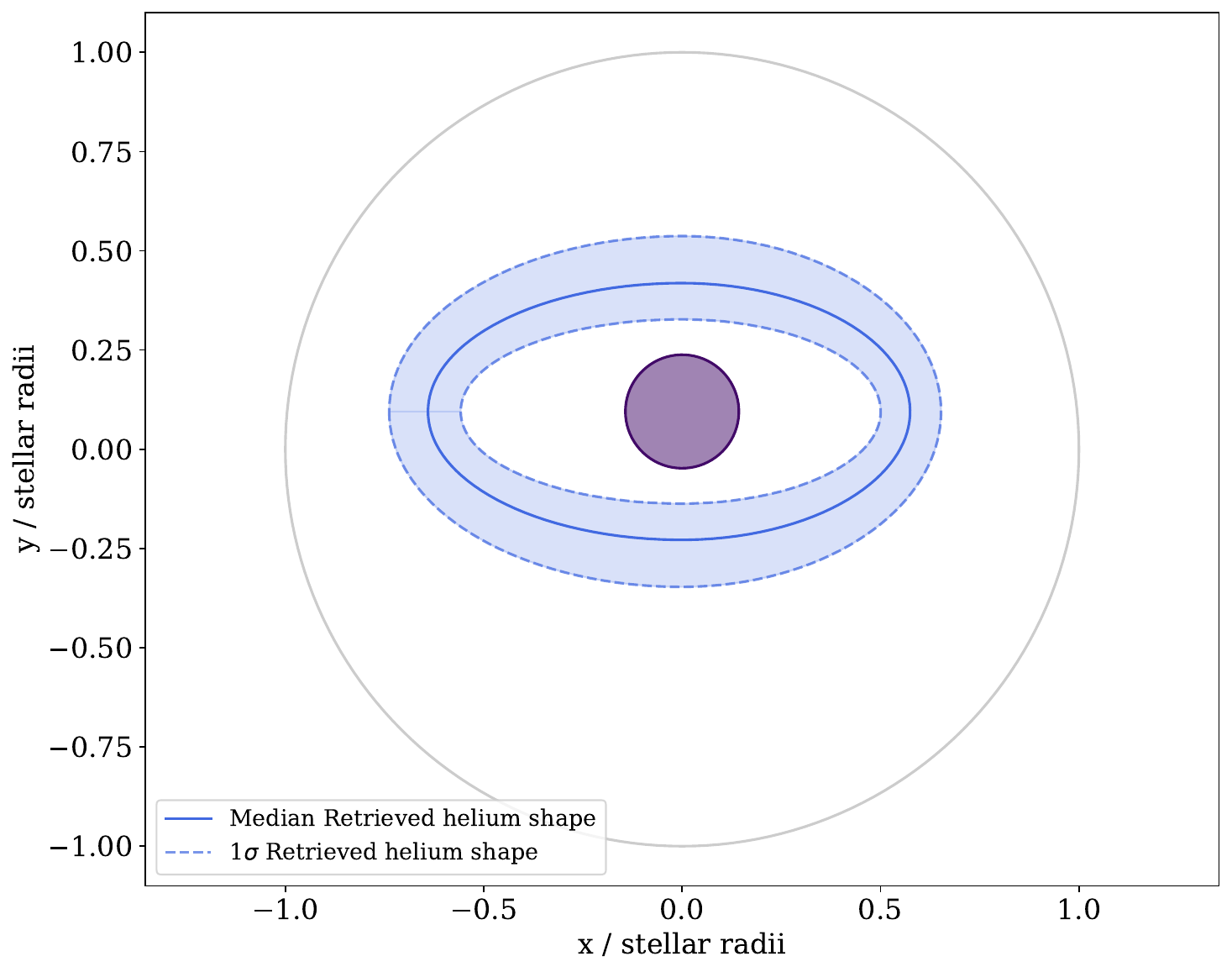}
    \caption{Main results of WASP-107b uniform helium envelope fit. The top left panel shows the Keck/NIRSPEC light curve in red, corresponding to a $\sim0.75$\AA~bin centered around the He triplet, together with the fitted model light curve in blue. The right panel shows the median retrieved envelope in blue, with the shaded region showing the $1\sigma$ uncertainty region. The gray line represents the star and the purple circle illustrates the planet.}
    \label{fig:w107}
\end{figure*}

We apply the proposed methodology to the ground-based observations of WASP-107b presented in \cite{Spake2021} and \cite{Kirk2020}. The observations were carried out with Keck/NIRSPEC and consist of two planetary transits observed in January 10, 2019 and April 6, 2019. As described in \cite{Spake2021}, we construct the light curve by combining the excess absorption  with the model white light curve presented in \cite{spake2018} (see Section 4.2 in \cite{Spake2021} for more details) . The corresponding helium light curve is shown in red in the left panel of Figure \ref{fig:w107}, compared to the model white light curve represented by the straight gray line.
Similarly to the analysis of HAT-P-18b, we fix the system parameters to the values given in \cite{spake2018}, and fit only for the four parameters describing the helium envelope. We perform the fits using the MCMC package \texttt{emcee}, running 20 walkers with 500 steps as burn-in and 2000 steps for the production run. We obtain a trailing and leading tail sizes of $l_1 = 4.48^{+0.68}_{-0.57} R_p$  and $l_2 = 4.02^{+0.55}_{-0.51} R_p$ respectively, together with $h = 2.26^{+0.83}_{-0.64} R_p$ an opacity of $1/\alpha = 2.90^{+1.40}_{-1.04}$. The results of this fits are plotted in blue in Figure \ref{fig:w107}, with the blue straight line in the left panel representing the fitted model light curve, and the right panel showing the corresponding helium envelope. The shaded regions represent  the $\pm1 \sigma$ uncertainty region. 
The results indicate that WASP-107b must posses large leading and trailing tails, which has been confirmed by recent JWST observations of the same target using NIRISS/SOSS \citep{krishnamurthy2025}. 

\begin{acknowledgments}
C.G. acknowledges funding from the Agency for Management of University and Research Grants from the Government of Catalonia (FI AGAUR) and from of a fellowship from ”la Caixa” Foundation (ID 100010434). The fellowship code is LCF/BQ/EU21/11890133. This work is part of the first author's doctoral thesis, within the framework of the PhD Program in Physics at the Universitat Autònoma de Barcelona.
M.M. gratefully acknowledges support from a Clay Postdoctoral Fellowship at the Smithsonian Astrophysical Observatory. H.R.W. and D.G. were funded by UK Research and Innovation (UKRI) under the UK government’s Horizon Europe funding guarantee as part of an ERC Starter Grant [grant number EP/Y006313/1]. We acknowledge support from Spanish grants PID2021-125627OB-C31 funded by MCIU/AEI/10.13039/501100011033 and by “ERDF A way of making Europe”, the programme Unidad de Excelencia María de Maeztu CEX2020-001058-M, and the Generalitat de Catalunya/CERCA programme. I.R. acknowledges financial support from the European Research Council (ERC) under the European Union’s Horizon Europe  programme (ERC Advanced Grant SPOTLESS; no. 101140786).
\end{acknowledgments}

The JWST data presented in this article were obtained from the Mikulski Archive for Space Telescopes (MAST) at the Space Telescope Science Institute. The specific observations analyzed can be accessed via \dataset[doi: 10.17909/1xak-8544]{https://doi.org/10.17909/1xak-8544}.

\bibliography{biblio}{}
\bibliographystyle{aasjournalv7}

\end{document}